\documentstyle[epsfig,sprocl]{article}
\def\be{\begin{equation}}
\def\bea{\begin{eqnarray}}
\def\eea{\end{eqnarray}}
\newcommand{\av}[1]{\mbox{$ \langle #1 \rangle $}}
\newcommand{\W}{\mbox{$W$}}

\newcommand{\xf}{\mbox{$x_F$}}  
\newcommand{\xp}{\mbox{$x_p$}}  
\newcommand{\xB}{\mbox{$x$}}  % Bjorken x
\newcommand{\Qsq}{\mbox{$Q^2$}}
\newcommand{\pz}{\mbox{$P_{\!z}^{\star}$}}
\newcommand{\pzb}{\mbox{$P_{\!z}^{'}$}}
\newcommand{\pzmax}{\mbox{$P^{\star}_{\!z{\tiny \rm max}}$}}
\newcommand{\pzmaxb}{\mbox{$P^{'}_{\!z{\tiny \rm max}}$}}
\newcommand{\pt}{\mbox{$P^{\star}_{\! T}$}}

\newcommand{\et}{\mbox{$E^{\star}_T$}}

\newcommand{\ptsq}{\mbox{$P_{\!T}^{\star 2}$}}

\newcommand{\gx}{\mbox{$g(x,Q^2)$}}

\newcommand{\lambdams}{\mbox{$\Lambda_{\rm \overline{MS}}$}}
\newcommand{\cm}{\mbox{\rm ~cm}}
\newcommand{\GeV}{\mbox{\rm ~GeV}}
\newcommand{\MeV}{\mbox{\rm ~MeV}}
\newcommand{\GeVsq}{\mbox{${\rm ~GeV}^2$}}

\newcommand{\jpsi}{\mbox{$J\!/\psi$}}
\newcommand{\sjpsi}{\mbox{\scriptsize $J\!/\psi$}}

\newcommand{\als}{$\alpha_s$}

\newcommand{\ep}{\mbox{$e^{\pm}p$}}
\newcommand{\mup}{\mbox{$\mu^{\pm}p$}}
\newcommand{\ee}{\mbox{$e^+e^-$}}
\newcommand{\pp}{\mbox{$p\bar{p}$}}
\newcommand{\fragfun}{\mbox{$1/\sigma_{\rm tot}\; d\sigma/dx_p$}}
\newcommand{\fx}{\mbox{$f(x,Q^2)$}}
\newcommand{\ftwo}{\mbox{$F_2$}}
\newcommand{\cms}{\mbox{cms}}
\newcommand{\Wgp}{\mbox{$W_{\! \gamma p}$}}
\newcommand{\Wgpsq}{\mbox{$W^2_{\! \gamma p}$}}
\newcommand{\Wgpd}{\mbox{$W^{\delta}_{\! \gamma p}$}}
\newcommand{\spsiel}{\mbox{$\sigma_{el}$}}
\begin{document}
%
%\mbox{}
% {\tt \vspace{-1.cm}   \hfill       H1-01/96-513 } \\
%\vspace{-0.2cm}
%\mbox{}
%
\title{Hadronic Final State in Deep-inelastic Scattering
       at HERA\footnote{
Invited talk given at XVI Int. Conf. on Physics in Collision,
Mexico City, June 1996.}}
\author{\vspace{-0.2cm}Tancredi Carli}
\address{
Max-Planck-Institut f\"ur Physik,
Werner-Heisenberg-Institut,
F\"ohringer Ring 6, \\
D-80805 M\"unchen,
Germany,
e-mail: h01rtc@rec06.desy.de \\
On behalf of the H1 and ZEUS collaborations.
}
%%%%%%%%%%%%%%%%%%%%%%%%%%%%%%%%%%%%%%%%%%%%%%%%%%%%%%%%%%%%%%
%
\maketitle\abstracts{
 Data on the hadronic final state of deep-inelastic
 events at the \ep~collider HERA are reviewed. 
 Fragmentation properties in the current region extracted
 from charged particle spectra are compared to \ee~and
 fixed target experiments at lower center of mass energies. 
 A measurement of the strong coupling constant from 
 integrated jet rates 
 is presented and prospects to extract the gluon 
 density in the proton by tagging vector mesons are discussed.
 Data on the inclusive mean transverse energy together with 
 the transverse momentum distribution of single charged
 particles indicate an increased gluon activity in the central
 rapidity region when the phase space
 is enlarged. The measurement of the forward jet cross-section is
 discussed.
 Keeping the phase space fixed, the mean transverse energy
 and the mean charged multiplicity for certain rapidity ranges is found 
 to behave analogously to \pp~collisions.
}
%
%
%\input jeth1fig.tex
%
%\vspace{-0.7cm}
\section*{Introduction}\label{sec:intro}
The large center of mass energy ($\sqrt{s} \approx 300$\GeV)
available at the \ep~collider HERA offers a large phase space
for production of hadrons in deep-inelastic scattering (DIS).
%Compared to previous DIS experiments, the invariant mass of
%all hadrons ($W$) is increased by a factor 10. 
This allows 
the detailed characteristics of the hadronic final state 
to be probed in a previously unexplored kinematic domain.

\begin{figure}
\vspace{-3.cm}
\begin{center}
\mbox{\hspace{2.cm}
\epsfig{figure=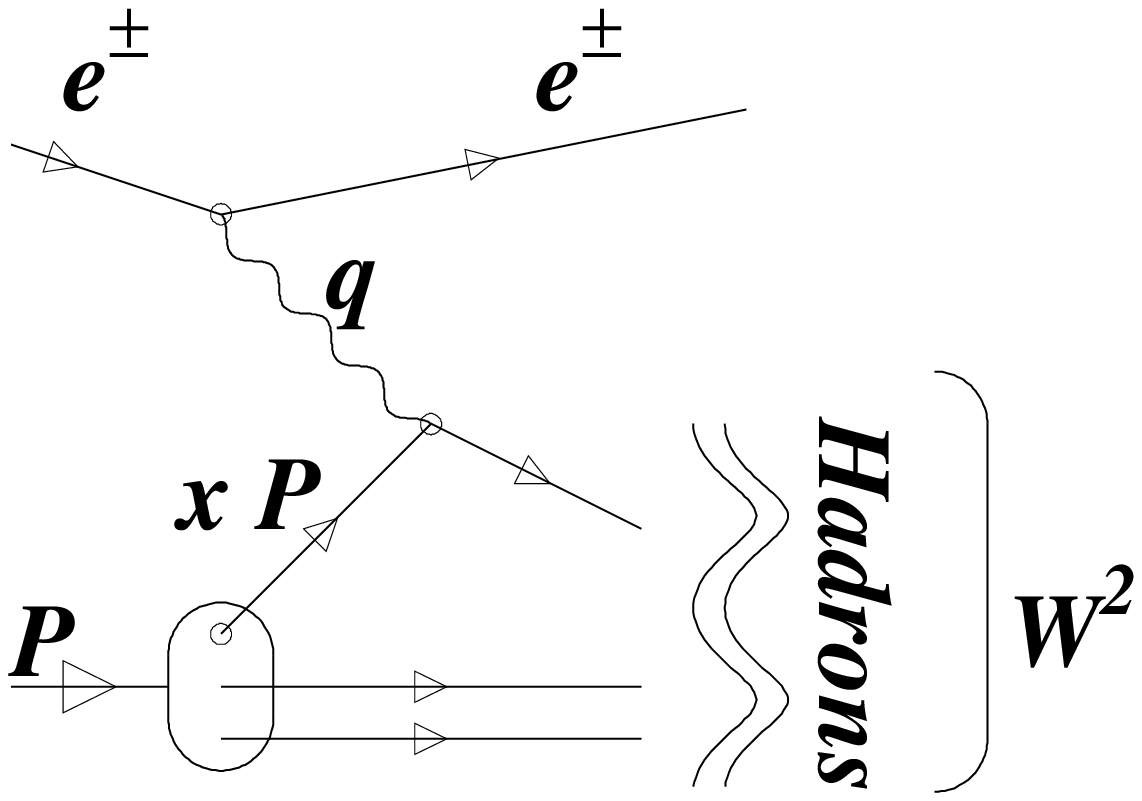,width=12.cm,height=9.cm}
}
\end{center}
\begin{picture}(0.,0.)
 \setlength{\unitlength}{0.8cm}
 {\Large
 \put(1.,5.) {$\sqrt{s} \approx 300$~{\rm GeV}}
 \put(11.,7.) {$x= \frac{Q^2}{p \cdot q } $}
 \put(11.,6.) {$W^2 \approx Q^2 \frac{1-x}{x}$}
 }
\end{picture}
\vspace{-2.8cm}
\caption{ Diagram of a simple electron-proton scattering at HERA.
\label{fig:qpm}}
\end{figure}

In the quark parton model (QPM) the photon scatters 
elastically on a quark freely moving in the proton. 
This is a good approximation when small distances are probed 
at large momentum transfers $Q^2$. The kinematics of the scattering
process is described
by the longitudinal momentum fraction of the quark with respect
to the proton momentum (\xB) and by the resolution of the photon
probing the proton structure ($Q^2$). These variables are defined
in a general way in Fig.\ref{fig:qpm} as Lorentz invariants.
At HERA, \xB~values down to a few $10^{-5}$ can be reached and
$Q^2$ ranges from low values where the photon is quasi on-shell
to very large ones resolving - with the present statistics - 
distances\cite{contact} down to $\approx 10^{-16}\cm$.

Perturbative QCD plays a key r\^ole
in the understanding of the complex mechanism governing the 
multi-particle final state.
For very short times (corresponding to short distances)
before and after the hard interaction of a virtual photon
with a parton in the proton, the partons are far away from their
mass shell and can be treated as quasi free particles. 
This first primary
scattering process can be perturbatively described by QCD.
Then long range forces become more and more important
and quarks with very low virtualities form bound states
leading to the observable hadrons.

%
%%%%%%%%%%%%%%%%%%%%%%%%%%%%%%%%%%%%%%%
%%%%%%%%%%%%%%%%%%%%%%%%%

The multi-purpose HERA detectors allow, in addition
to the scattered electron, a large part of the hadronic
final state to be measured. 
Provided that the transition from partons to 
hadrons can be properly modeled, this opens the possibility to
study the underlying parton dynamics using exclusive observables 
built on particle spectra, jets or energy depositions.
Variables based on the hadronic final state can be
defined for specific rapidity ranges such that e.g. properties
of hadrons close to the proton remnant can be investigated.
It is also possible to isolate the current region where the hard
subprocess induced by the photon virtuality has taken place
and to compare with results obtained in \ee annihilation.
An advantage of HERA is that most measurements can be performed 
at a variable scale.
The mean multiplicity of charged particles can e.g. be studied as
function of $W$ and the strong coupling constant \als~ can be
extracted from integrated jet rates in different bins of $Q^2$.
The phenomenon of diffraction in DIS is covered by M.~Derrick 
in these proceedings.

\section*{ Particle spectra in the current region}\label{subsec:spectra}
A natural reference frame to study properties of the hadronic final
state is the hadronic center of mass system (\cms) where
the virtual photon and the proton have momenta of equal
magnitude and point in opposite directions.
Within the QPM, the photon and the incoming quark collide
in this frame head-on and no transverse momentum (\pt) is
transfered to the struck quark. The \pt~of the measured hadrons is only
produced during the fragmentation of the quark and therefore maximally
limited to the typical inverse size of a hadron. 
Higher values of \pt~can only originate from
gluon emissions before and after the hard interaction.
Neglecting gluon radiation
the longitudinal momentum (\pz) of the struck quark is 
given by $W/2$. Correspondingly the target remnant moves in
opposite direction with $-W/2$.% (see Fig.~\ref{fig:cms}).
A natural variable to study the longitudinal
momenta of the emerging hadrons is therefore
$\xf = \pz/\pzmax = 2 \pz/W$.

\begin{figure}
\vspace{-0.15cm} 
\begin{center}
%\setlength{\unitlength}{0.8cm}
% \put(0.,-0.5) {(a)}
% \put(6.,-0.5) {(b)}
 \epsfig{figure=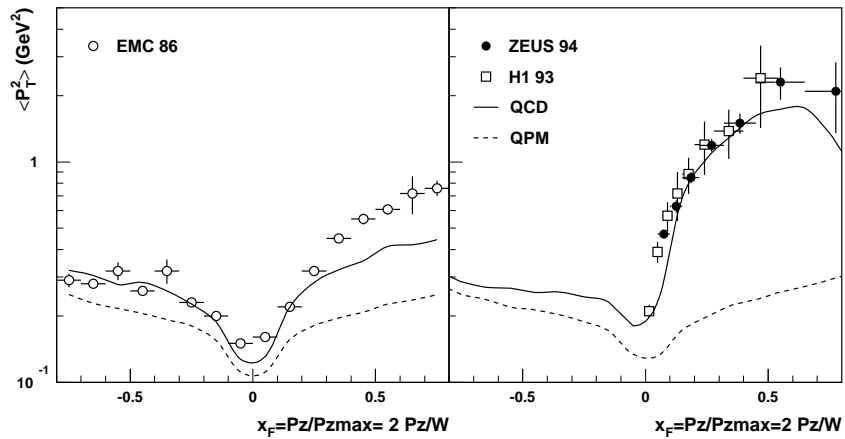,width=12.cm}
\end{center}
\vspace{-0.2cm} 
\caption{\it Mean squared transverse momentum as function of the
scaled longitudinal momentum for charged
particles in the cms. HERA data are indicated by the open squares (H1)
and the closed circles (ZEUS). The data from a \mup~fixed target
experiment at $\sqrt{s}=23$\GeV~(EMC) are given
as open circles. The prediction from ARIADNE as a QCD model (solid) and the
naive quark parton model (dashed) are superimposed as lines. 
\label{fig:seagull}}
\vspace{-0.3cm} 
\end{figure}

The importance of processes involving QCD radiation is illustrated
in Fig.~\ref{fig:seagull} where the 
\av{\ptsq}~versus the \xf~of charged particles is shown.
Without allowing for parton radiation the \av{\ptsq}
is limited to $0.3$\GeVsq~(dotted line). Much larger 
\av{\ptsq}~are found in the HERA data~\cite{h1eflow93,zeussg} 
and they are well described by a QCD model (solid line).
In the target hemisphere ($\xf<0$), where no HERA measurements
are available yet, the \av{\ptsq} is less prominent and
only little \pt~is produced by gluon radiation.
%This indicates the considerable impact of non-perturbative processes
%on the proton remnant.
This can be understood in terms 
of the extended colour charge distribution supressing small wavelength
gluon radiation.
At lower center of mass energies where less phase space
for gluon radiation is available, the \av{\ptsq} is even more
limited. This can be seen e.g. in the data of the EMC 
collaboration\cite{emc86},
a \mup~fixed target experiment at $\sqrt{s}=23$\GeV. 
In this kinematic region QCD models 
do not significantly add transverse momenta to the
naive expectation from the QPM and have moreover difficulties to 
describe the data~\cite{tuning91}. 
%%%%%%%%%%%%%%%%%%%%%%%%%

\begin{figure}
%\rule{5cm}{0.2mm}\hfill\rule{5cm}{0.2mm}
\vspace{-0.3cm}
%\rule{5cm}{0.2mm}\hfill\rule{5cm}{0.2mm}
%
\begin{center}
 \begin{tabular}{cc}
\setlength{\unitlength}{0.8cm}
 \put(0.,-0.5) {(a)}
 \put(6.,-0.5) {(b)}
 \mbox{ \hspace{-2.cm}
 \epsfig{figure=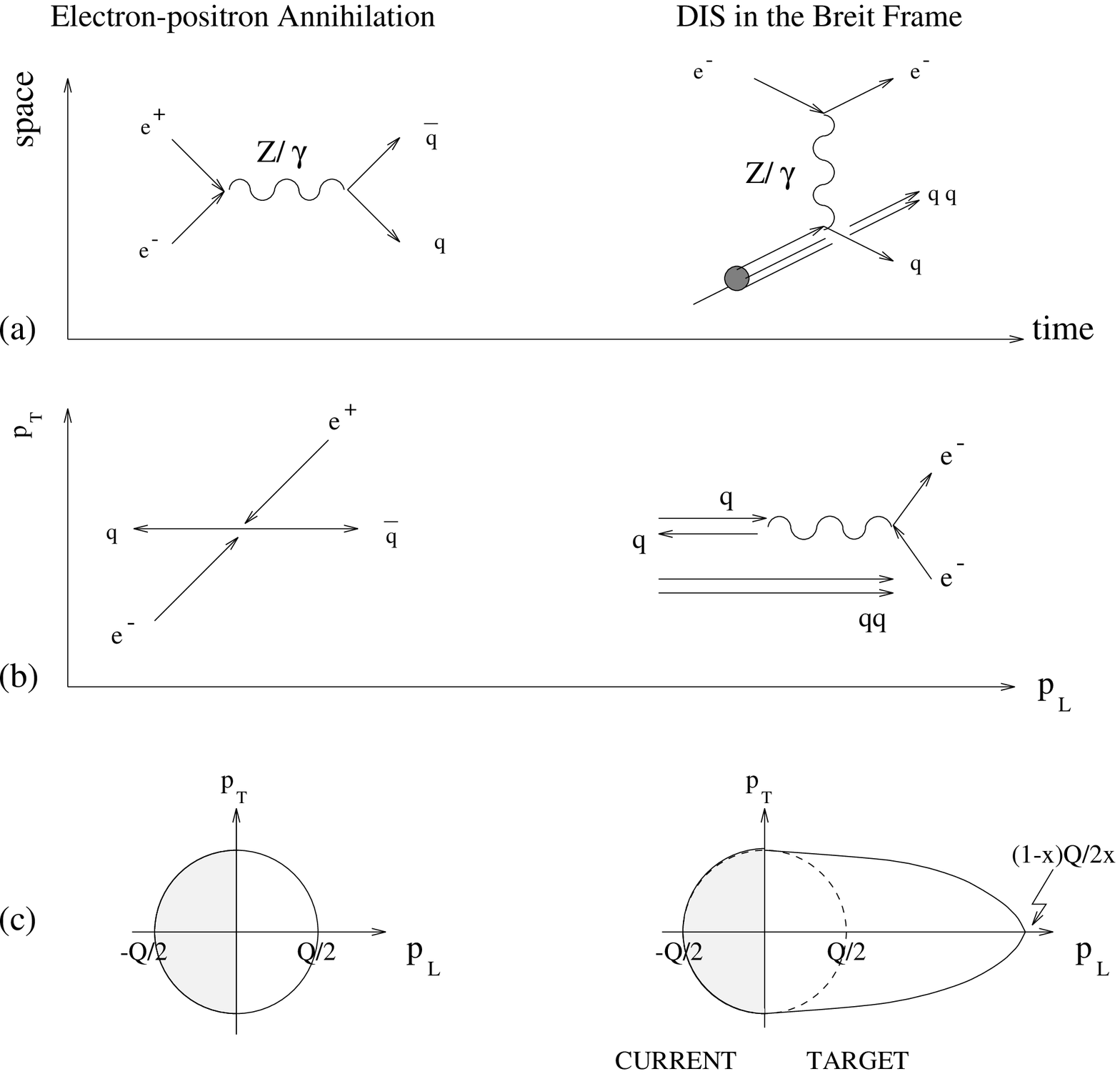,bbllx=211,bblly=300,bburx=335,bbury=513,
         clip=,angle=270,width=6cm}
 \epsfig{figure=breit.eps,bbllx=390,bblly=316,bburx=555,bbury=583,
         clip=,angle=270,width=6cm}
 }
\end{tabular}
 \end{center}
\caption{\it
(a) Sketch of a simple photon-quark collision in the Breit frame 
(b) Phase space for the longitudinal and transverse momenta
    of single particles in this frame.
\label{fig:breit}}
\vspace{-0.3cm}
\end{figure}

Detailed comparisions between \ee~and \ep~collisions are most suitably
carried out in the Breit frame.
In this frame the only non-zero space-like component of  the 
photon four-momentum is given by $\pzb=-2 \xB E_P$
with $E_P$ the proton beam energy.
Within the QPM, the struck quark enters with $\pzb=Q/2$ and collides on
the virtual photon as on a 'brick wall', from which it simply rebounds
with $\pzb=-Q/2$ (see Fig.~\ref{fig:breit}a). 
The particles scattered in the hemisphere of the struck
quark ($\pzb<0$) behave like those produced in one hemisphere of an 
\ee~annihilation event at $s=$\Qsq. As in \ee~in this hemisphere properties 
of longitudinal particle momenta should only depend on $Q$.
The opposite hemisphere ($\pzb>0$) contains the proton remnant
and moves with $\pzb=(1-\xB)E_P=(1-\xB)Q/2\xB$ (see Fig.~\ref{fig:breit}b).
%%%%%%%%%%%%%%%%%%%%%%%%%%%%%%%%%%%%%%%
%%%%%%%%%%%%%%%%%%%%%%%%%

%\input blabla
%%%%%%%%%%%%%%%%%%%%%%%%%%%%%%%%%%%%%%%
%\vspace{-9.cm}
\begin{figure}
\vspace{-0.2cm}
\begin{center}
%\mbox{\hspace{2.cm}
\epsfig{figure=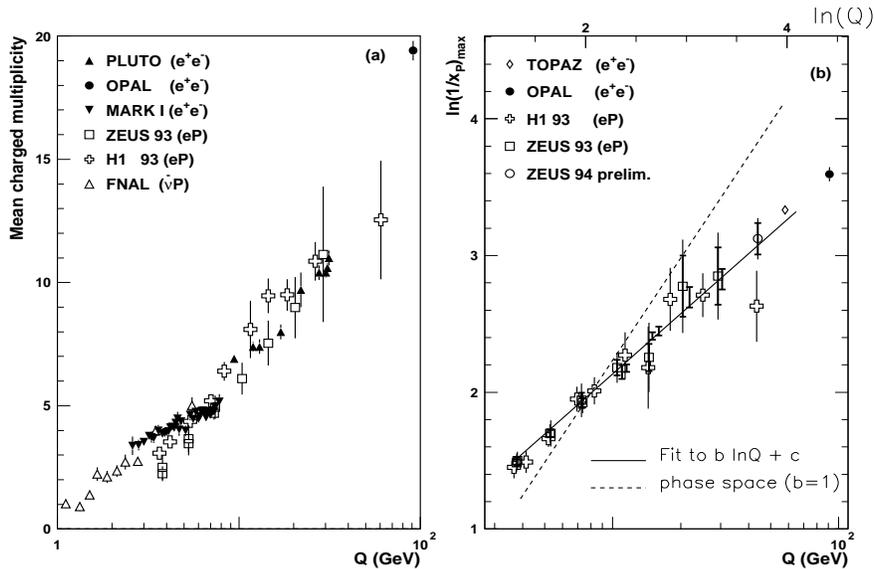,width=12cm,height=8.cm}
%}
\end{center}
%\begin{picture}(0.,0.)
% \setlength{\unitlength}{0.8cm}
% {\Large
%% \put(1.,5.5) {$\sqrt{s} \approx 300$~{\rm GeV}}
% \put(10.,6.8) {$x= \frac{Q^2}{p \cdot \gamma} $}
% \put(10.,5.2) {$W^2 \approx Q^2 \frac{1-x}{x}$}
% }
%\end{picture}
\vspace{-0.8cm}
\caption{\it Mean multiplicitiy of charged particles (a) 
and the peak position of the $\ln{1/\xp}$ distribution (b) as function of $Q$
in \ee, \ep~and $\bar{\nu}p$ collisions.
The \ep~data are measured in the Breit frame and the multiplicity is multiplied
by a factor of two. 
\label{fig:coherence}}
\vspace{-0.3cm}
\end{figure}
%%%%%%%%%%%%%%%%%%%%%%%%%

The linear rise of (twice) the mean charged particle multiplicity
(\av{$n$}) with $\log{Q}$ for DIS measured at HERA\cite{nvsq2h1,nvsq2zeus}
is shown in Fig.~\ref{fig:coherence}a.
For $Q>7$\GeV, DIS and \ee~data\cite{nvsq2ee} reasonably agree.
However, when going to lower $Q$ the differences between
\ee~and \ep~collisions become more and more distinct.
A higher multiplicity is observed in \ee~data taken by MARK1.
A clear dependence of \av{$n$} on \xB~is observed when comparing
the HERA data at low \xB~with the $\nu p$
data of a fixed target experiment at FNAL\cite{fnal} at \xB~close to
$0.1$.
While agreeing at high \xB~with the \ee~data, in DIS the current
hemisphere is more and more depopulated at low \xB~where higher order
parton emission become increasingly important.
This is due to an asymmetric migration from the current to the target
hemisphere, not present in \ee~annihilation.

\begin{figure}
%\vspace{-0.2cm}
\epsfig{figure=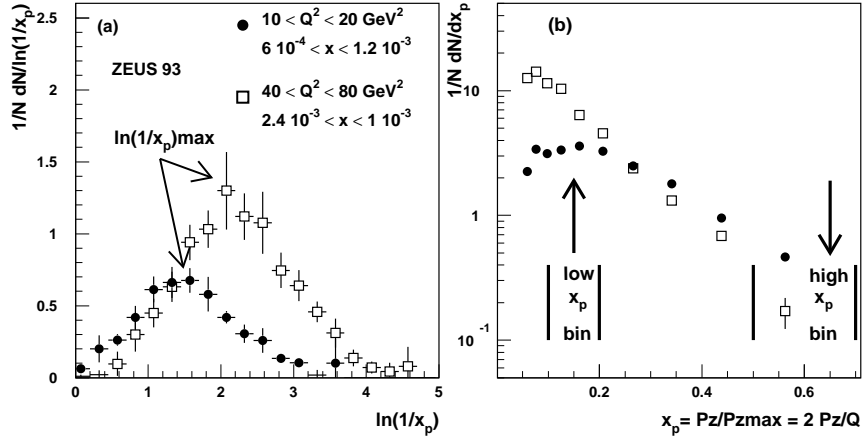,width=12.cm}
%\vspace{-0.5cm}
\caption{\it
 Distribution of the scaled longitudinal
 momentum $\xp=2 \pzb/Q$~for charged particles in the Breit
 frame in two $(\xB,\Qsq)$ bins measured by {\rm ZEUS} (b). 
 The same data after transforming \xp~to $\ln{1/\xp}$ (a).
\label{fig:xp}}
%\vspace{-1.2cm}
%
\end{figure}

The distribution of the scaled longitudinal momenta
$x_p=\pzb/\pzmaxb= 2\pzb/Q$ of single charged particles
allows fragmentation to be studied in further detail.
The fragmentation function \fragfun, with
$1/\sigma_{\rm tot}$ the total cross-section in
a particular (\xB,\Qsq) bin, characterizes the complete
process of hadron formation including QCD parton evolution and
the non-perturbative transition from partons to hadrons.
It is not calculable in perturbative QCD, but can, as the structure
function \ftwo, be evolved in \Qsq~given a measured starting
distribution.
\fragfun~rises with decreasing \xp~and turns over at low \xp.
This region can be expanded by transforming
\xp~to $\ln{1/x_p}$ (see Fig.\ref{fig:xp}a and b).
At small \xp, \fragfun~is affected by coherence of soft gluons
while at large \xp~it can be used to determine \als.

At small \xp~the phase space for soft gluon emission is reduced to an
angular ordered region caused by destructive interference.
This suppression of soft gluons is predicted by QCD, since the power of
gluons to resolve individual colour charges should be diluted with
increasing wavelength\cite{dokshitzer}. Assuming that the basic
features of a distribution for partons are preserved for the measured hadrons,
this results then in a slower rise of the
particle multiplicity and of the peak position of the $\ln{1/x_p}$
distribution with $\log{Q}$ than simply expected from the increased 
longitudinal phase space. 
Fig.~\ref{fig:coherence}b shows the evolution of the
$\ln{1/x_p}$ peak position measured for particles having $\xp > 0$.
The same linear rise with $\log{Q}$ is observed in \ee~data\cite{ln1xpee}
which provides a hint of a universal behaviour of quark fragmentation.
As the data become more precise, expected differences between
\ee~and DIS fragmentation should become manifest due to e.g. different
quark flavour composition.
%%%%%%%%%%%%%%%%%%%%%%%%%%%%%%
%%%%%%%%%%%%%%%%%%%%%%%%%

The modified leading logarithm approximation (MLLA) of QCD is able to
describe this evolution down to very low $Q$ (solid line).
% important test of QCD   % (recently also pp Wo ?) % NLO
The coherent evolution is clearly favoured by the data.
It can, however, not be excluded that a similar attenuation of
the slope could also be caused by effects during the
hadronisation phase\cite{h1breit}.
%%%%%%%%%%%%%%%%%%%%%%%%%%%%%%
\begin{figure}
%\begin{wrapfigure}{r}{8cm}
\vspace{-2.cm}
\begin{minipage}[b]{12.cm}
 \hspace{-1.cm}
 \begin{minipage}[t]{8.cm}
  \makebox[0cm]{}
  \epsfig{figure=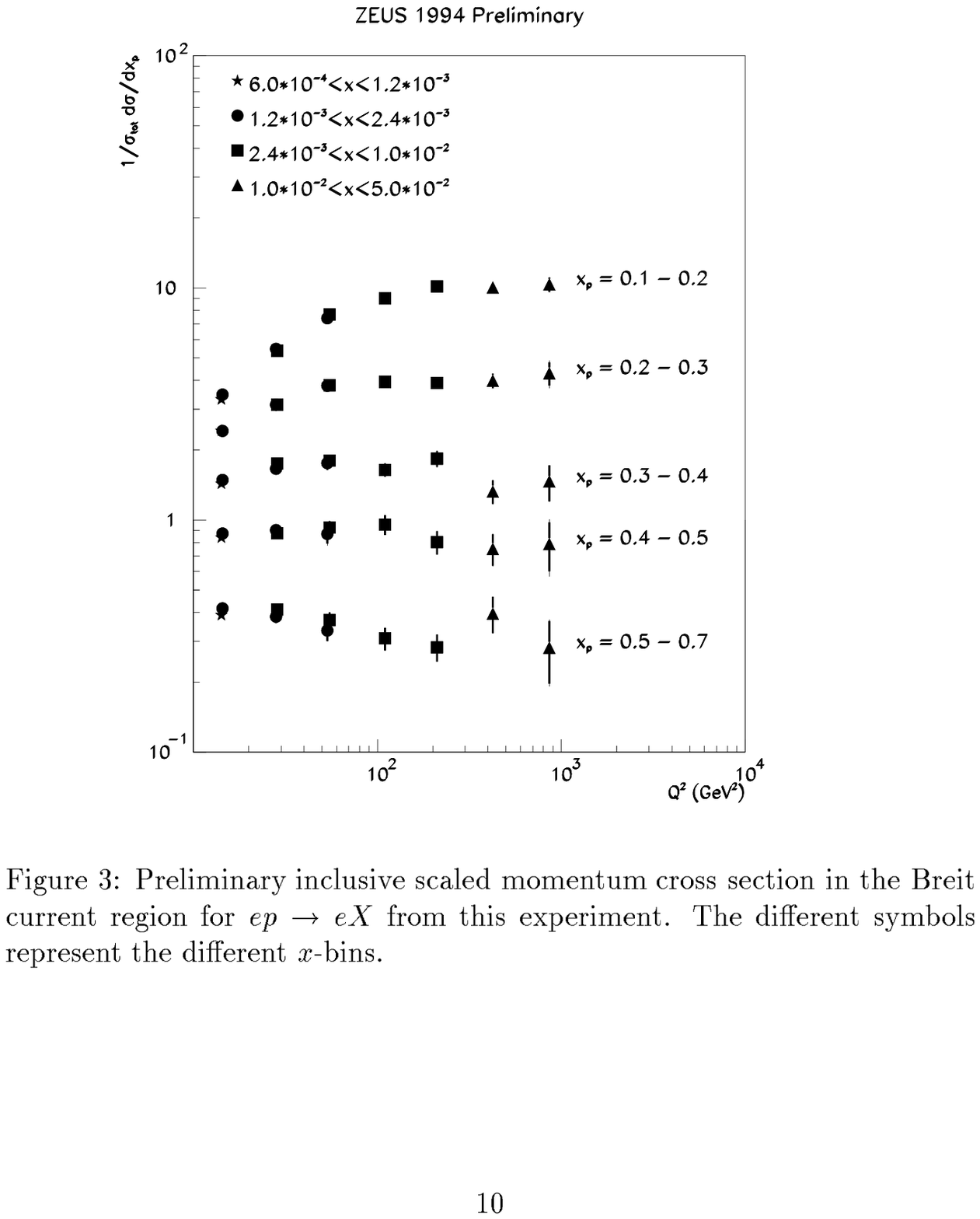,width=8.cm,
         bbllx=113,bblly=643,bburx=422,bbury=302,clip=}
 \end{minipage} \hfill
% \hspace{8.cm}
%\vspace{1.cm}
 \parbox[t]{4.cm}
 {\makebox[0cm]{}
  {\it \hfill \\  \hfill \\  \hfill \\  
        \hfill \\  \hfill \\  \hfill \\  
      \caption{}
      \fragfun in the Breit frame
       for different bins of the scaled longitudinal
       momentum \xp~as function of \Qsq. The data have been
       measured by {\rm ZEUS}.  \label{fig:fragfun}
  }
 }
\end{minipage} \hfill
%
% \vspace{8.2cm}
\vspace{3.cm}
\end{figure}
%\end{wrapfigure}
%%%%%%%%%%%%%%%%%%%%%%%%%

The evolution of \fragfun~in different bins of \xp~with
\Qsq~is shown in Fig.~\ref{fig:fragfun}\cite{zeusfragfun}.
With increasing \Qsq, \fragfun~ increases at low \xp~and 
decreases at high \xp.
When more phase space is available more gluons can be
emitted which results in a softening of the particle spectra.
Therefore more particles at low \xp~and less at large \xp~are observed.
These scaling violations can be treated in an analogous way
as in the proton structure function analysis\cite{ball} to extract the
strong coupling constant \als. Such a measurement has been
performed~\cite{lepalphas} using LEP data together with 
lower energy \ee~data.
At HERA this method could be applied in one single experiment.
However, to avoid the region were migrations between the current and
target hemisphere become important, one has to restrict
the analysis to \Qsq$>100$\GeVsq.
Scaling violations are observed, but more data is needed to get
the necessary precision at large \xB~and \Qsq~for
a competitive extraction of \als.
Since the scaling violation in the fragmentation function are
only of logarithmic nature, the achievable error on \als~will,
even with high statistics, not be competitive with the present error
on the world average\cite{grauscal}.
\section*{Determination of the strong coupling constant
%and the gluon density
using jets}
\label{sec:alphas}
\begin{figure}
%\rule{5cm}{0.2mm}\hfill\rule{5cm}{0.2mm}
%\vskip 2.5cm
%\rule{5cm}{0.2mm}\hfill\rule{5cm}{0.2mm}
%\hspace{-1cm}
\epsfig{figure=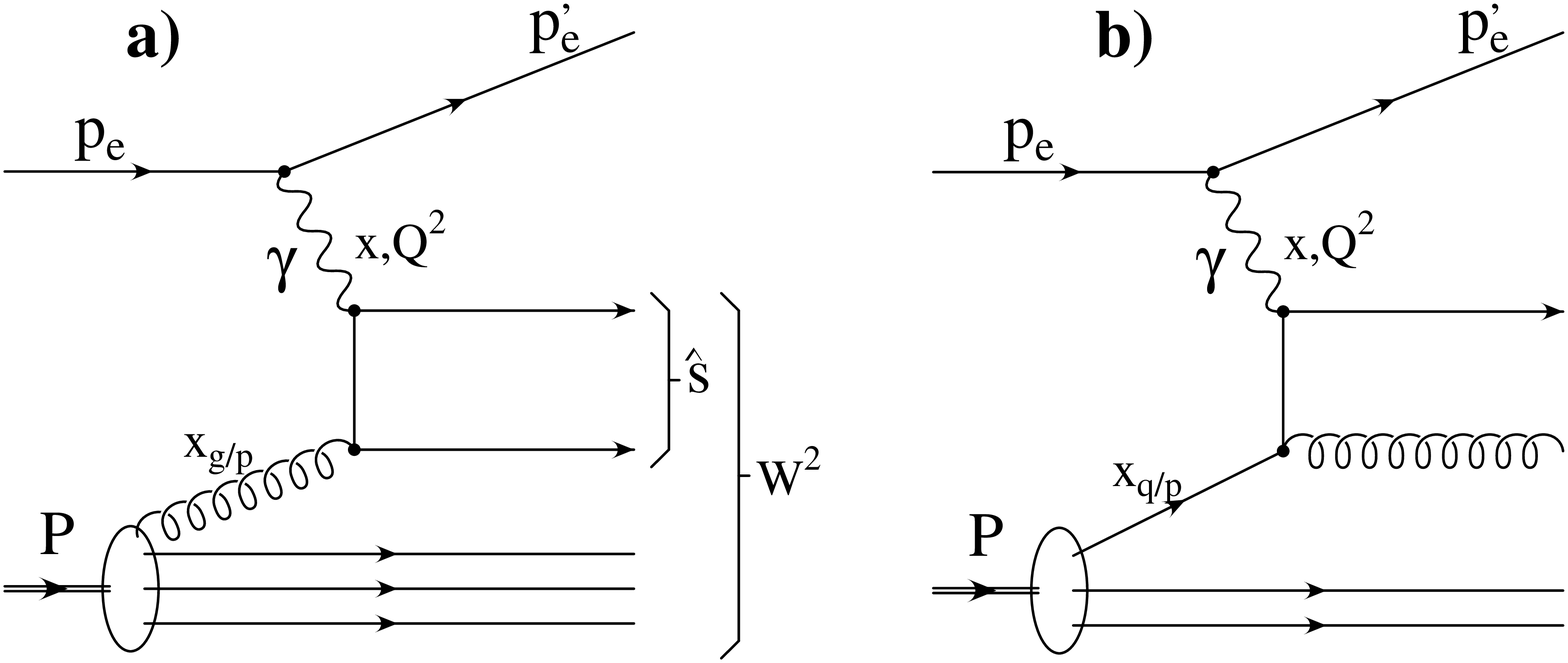,width=10cm}
%        bbllx=401,bblly=729,bburx=137,bbury=323
%        ,width=8cm,angle=270,clip=}
\caption{Feynman diagrams for the production of $2+1$~jet events 
to first order of \als~ in \ep-collisions.
\label{fig:feynjets}}
\vspace{-0.5cm}
\end{figure}

The high values of $W$ reachable at HERA allow 
clean jets to be observed in the hadronic final state.
Events with two jets in addition to the one close to the beam
carrying the remnants of the proton (2+1 jet events)
are produced in leading order of \als~by processes 
either induced by a gluon in the proton (Fig.~\ref{fig:feynjets}a)
or by a quark radiating a gluon before or after interacting with the
photon (Fig.~\ref{fig:feynjets}b). 
The cross-section of both processes is proportional to \als.
For the gluon initiated processes
in addition precise knowledge of the gluon density in the proton 
\gx~is required to calculate the cross-section.
%%%%%%%%%%%%%%%%%%%%%%%%%%%%%%
%%%%%%%%%%%%%%%%%%%%%%%%%

The sensitivity to both \als~and \gx~can be used
to simultaneously extract these quantities. Such an analysis is however
theoretically and experimentally difficult. The analysis is therefore
split in two kinematic regions.
At low \xB~and \Qsq~the gluon initiated processes dominate.
This can be used to extract \gx~by assuming
a value of \als~and statistically subtracting the quark induced processes.
At large \xB~and \Qsq~the parton densities are well known and
the proportionality coefficients between the 2+1~jet rate and
\als~are given by next-to-leading order QCD 
calculations\cite{nloqcdjet}. 
The measurement of the ratio of 2+1~jet events
to all events at some value of $Q$ allows therefore 
$\alpha_s(Q)$ to be extracted. 
By evaluating $\alpha_s(Q)$ in different $Q$~bins
the running of the strong coupling constant can be tested
in one experiment.

\begin{figure}
\vspace{-0.2cm} 
\begin{center}
%\setlength{\unitlength}{0.8cm}
% \put(0.,-0.5) {(a)}
% \put(6.,-0.5) {(b)}
 \mbox{\hspace{-1.cm}
 \epsfig{figure=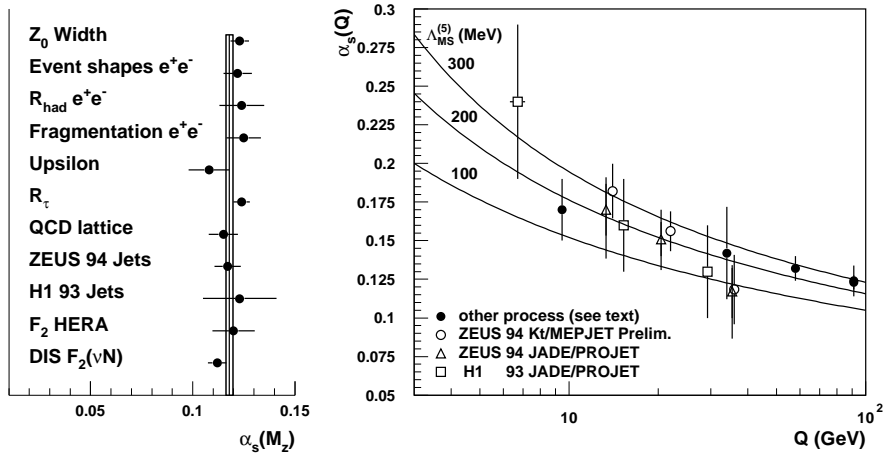,width=13.cm}
}
\end{center}
\vspace{-0.2cm} 
\caption{Left: Values and total error of $\alpha_s(M_Z)$ from 
 various processes. 
 The solid line indicates the world average %\cite{pdg} 
 and the band its total error. 
 Right: \als($Q$) from HERA (open symbols) and other processes
 with increasing $Q$ (closed circles): 
% decay width of the upsilon meson, 
 $\Gamma_{\! \Upsilon}$ and
 $\sigma_{\rm \! had}/\sigma_{\rm \! tot}$,
 event shapes and $\Gamma_{\rm \! hadron}/\Gamma_{\rm \! lepton}$ 
 in \ee.
 The predictions of the renormalisation group
 equation for three values of \lambdams~ are superimposed as lines. 
\label{fig:alphas}}
\vspace{-0.5cm} 
\end{figure}

For the \als~measurement only the kinematic range of high $Q$~is
considered to reduce the uncertainties from hadronisation. 
To avoid the region where higher order gluon emission
become important, only events at high \xB~are
% (e.g. by an explicit cut in \xB$>0.01$) 
considered, and also, the
experimentally difficult, very forward region is excluded.
%demanding e.g. $z_p >0.1$. The invariant $z_p$ is, with $p_{\rm jet}$ 
%the four-momentum of one of the two jets, defined by
%$z_p= P \cdot p_{\rm jet}/P \cdot \gamma$ and corresponds in the photon-parton
%center of mass system to $1/2 \; (1-\cos{\theta^*_{\rm jet}})$.
Jets are experimentally and theoretically defined by the JADE clustering
algorithm\cite{jade}.%, where a distance between particles is defined by
%$y_{ij}=m_{ij}/W^2$ ($m_{ij}$ the invariant mass of a particle pair) and pairs
%with smallest $y_{ij}$ are clustered together until $y_{ij}$ are above a fixed
%resolution parameter $y_{\rm cut}$. The proton remnant is included
%in this procedure as a pseudo-particle carrying all the longitudinal
%momentum lost in the beam pipe.
%
%For  $y_{\rm cut}=0.2$, 
Three values of \als($Q$) extracted
by the H1~\cite{h1alphas} and ZEUS~\cite{zeusalphas}
collaboration %with the JADE algorithm 
are shown in Fig.~\ref{fig:alphas}. For increasing values of
$Q$, \als($Q$) decreases as predicted by the renormalisation
group equation shown for \lambdams~between $100$ and
$300$\MeV~(solid lines).
An extrapolation to \als($M_z$) yields:
\begin{eqnarray}
\vspace{-0.1cm}
\label{eq:alphas}
 \nonumber \! \! \! \! \! \! \! \! \! \! 
 {\rm H1 \; 93:  } \; \; \; \; \; \; \; \; \; \; 
  \alpha_s(M_z)  = 0.123 \pm 0.012 {\rm(stat)}
 \pm 0.013 {\rm (syst.)} \\
 \nonumber
 {\rm ZEUS \; 94:} \; \; \alpha_s(M_z) = 0.117
 \pm 0.005 {\rm (stat)} + 0.005 {\rm(exp.)} \pm 0.007 {\rm (th.)}
\end{eqnarray}
%
%%%%%%%%%%%%%%%%%%%%%%%%%%%%%%%%%%%%%%%%%%%%%%%%%%%%%%%%
%%%%%%%%%%%%%%%%%%%%%%%%%%%%%%%%%%%%%%%%%%%%%%%%%%%%%%%

%
% combining\cite{pdg} the two values leads to 
% alpha_s=0.121\pm 0.004(stat)\pm 0.008 (syst)
which is consistent with other values obtained from
a large variety of different processes as shown Fig.~\ref{fig:alphas}
(see\cite{pdg} for references).
The agreement found in \ee, \ep~and \pp~collision for reactions
at very different scales is an important and successful test of QCD.
The error on the HERA measurement are even with the present
limited statistics already competitive.

The dominant experimental errors stem from the uncertainty on the
energy scale for hadrons, the model dependence of assigning
the jets to partons and the phenomenological description of
the hadronisation process. Furthermore the renormalisation scale
and the choice of the input parton density systematically influences
the result. For consistency it would be necessary to
simultaneously fit \als~used in the parton densities, but this effect
is negligible in the phase space region where
the HERA analysis has been performed\cite{chyla}.
Recently a new next-to-leading order calculation\cite{mepjet}
became available allowing to analyze the data using all available jet 
definition schemes
and to impose any acceptance cut on the final state particles.
Moreover different renormalisation and factorisation scales
(e.g. \pt~instead of $Q$) can be tested.
These calculations will allow 
the theoretical systematic errors to be better assessed.
One of the insights gained from these studies is that the
cone\cite{conealgo} or $k_T$\cite{ktalgo} jet algorithm seem to
be better suited for precision QCD tests than the JADE scheme,
since the variation of the cross section when going from leading
to next-to-leading order is much smaller\cite{mirkes}. Moreover, these jet
algorithms are less dependent on the choice of the hard scale
and allow a larger phase space to be covered.
%       
%The $k_T$ algorithm groups objects like e.g. particles
%in the Breit frame to jets. For each pair of particles
%$y_{ij}=2 {\rm min}(E_i^2,E_j^2) (1 - \cos{\theta_{ij}})/Q^2
%\approx  {{k_T}_{ij}}^2/Q^2$ where
%$k_t$~represents the transverse momentum of the particle $i$ relative to
%$j$ or vice versa. If $y_{ij}$ is smaller then a resolution parameter
%$y_{\rm cut}$, the two particles are merged.
%A fixed momentum $r$ is introduced to account for the
%proton remnant and $y_{ir}=2 E^2_i (1-\cos{\theta_{ir}})/Q^2$ is computed
%along with $y_{ij}$ at each stage.
%If the smallest of all $y_{ij}$ is $y_{kr}$, then particle $k$
%is classified as part of the remnant and is not available for further
%clustering.
% An additional advantage of the $k_T$ algorithm is that
% the difference between the leading and next-to-leading order are smaller
% the influence of higher order gluon emission is relatively small
% and no cut in $z_p$ is needed to avoid the forward region.
% Hence, the covered phase space is larger.
%
 The ZEUS collaboration has reanalyzed their data\cite{trefzger} using the
 $k_T$ algorithm. The preliminary values of \als($Q$) obtained
 in three bins of $Q$ are shown\footnote{Only the statistical
 errors are included in the figure.} in Fig.~\ref{fig:alphas}.
 They are consistent with the results obtained
 with the JADE algorithm.
\section*{On the determination of the gluon density in the proton}
 The dynamics of the strong interaction at high momentum scales,
 corresponding to short distances, is successfully described
 by perturbative QCD. A vital ingredient of any QCD prediction
 for a reaction where a nucleon is involved is an assumption about
 the parton distribution in the nucleon needed to calculate an
 observable in terms of basic QCD subprocesses at the partonic level.
 This parton density function \fx~is the probability to find a parton within
 a nucleon carrying a fraction \xB~of its momentum when probed
 by a particle with virtuality \Qsq. Once \fx~has been determined
 in one reaction it can be used for any other prediction.

 In deep-inelastic scattering the collinear singularities encountered when
 calculating the hard subprocess to order \als~are absorbed
 in the parton densities. The parton densities
 are usually evolved by the DGLAP equations\cite{dglap},
 presently known in leading and next-to-leading order,
 which describe the splitting of a parton.
 Once the parton densities have been measured at some
 \Qsq~value they can be evolved to any point in the kinematic plane.
 The parametrisation of the input distribution is obtained
 by a global fit to a large variety of data. Two different
 strategies have been adopted.
 Martin, Roberts and Stirling\cite{mrsh} and the 
 CTEQ collaboration\cite{cteq}
 parametrise the starting distribution at some
 $Q^2_0$ ($1-5$\GeVsq) and fit then the parameters to 
 data by evolving $f(x,Q^2_0)$.
 A dynamical approach is pursued by Gl\"uck, Reya and Vogt\cite{grv}
 where only valence like parton densities with $f(x=0,Q^2_0)=0$
 are assumed at a very low input scale $Q^2_0=0.34$\GeVsq.
 The gluon and the sea quark distributions are then generated
 radiatively. This procedure %was the only one which %technique
 predicted the steep rise of the structure function~\footnote{see 
 p.14-15 for more details and references.} \ftwo~with
 decreasing \xB~and moreover the slowing down of this rise towards 
 lower \Qsq.
 %The original intention: input valence
 %quarks would suffice, gluon and sea distribution generated
 %radiaitvley However, sizeable valence gluon and valences sea distribution
 %are also required at the input scale to describe data
 %GRV found in good agreement with HERA data down to very low values Q2=1.5 GEV

%%%%%%%%%%%%%%%%%%%%%%%%%%%%%%%%%%%%%%%
\begin{figure}
\vspace{-0.3cm}
%\rule{5cm}{0.2mm}\hfill\rule{5cm}{0.2mm}
%\vskip 2.5cm
%\rule{5cm}{0.2mm}\hfill\rule{5cm}{0.2mm}
%
\begin{center}
 \begin{tabular}{cc}
  \epsfig{figure=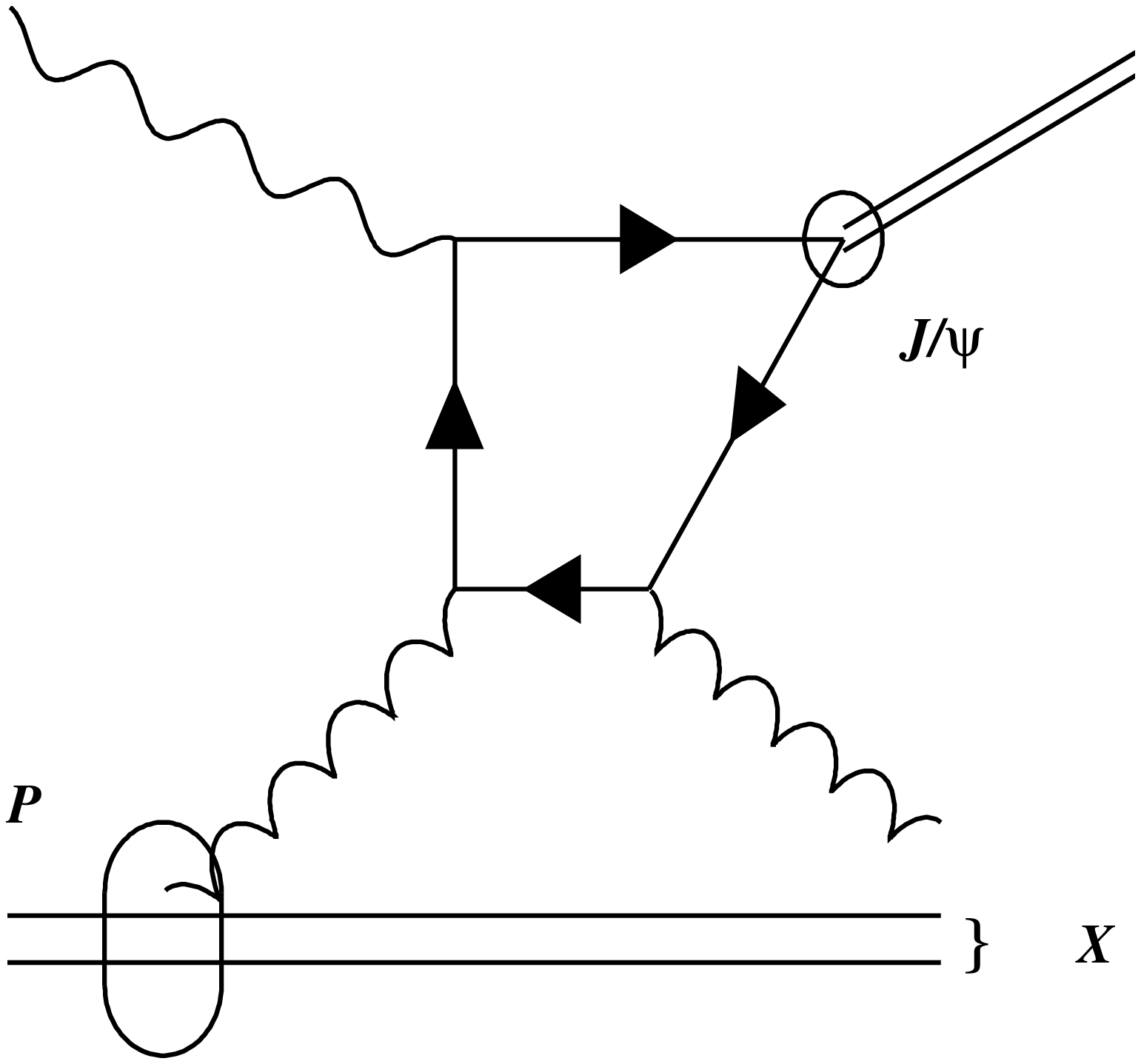,width=6.cm}
  \epsfig{figure=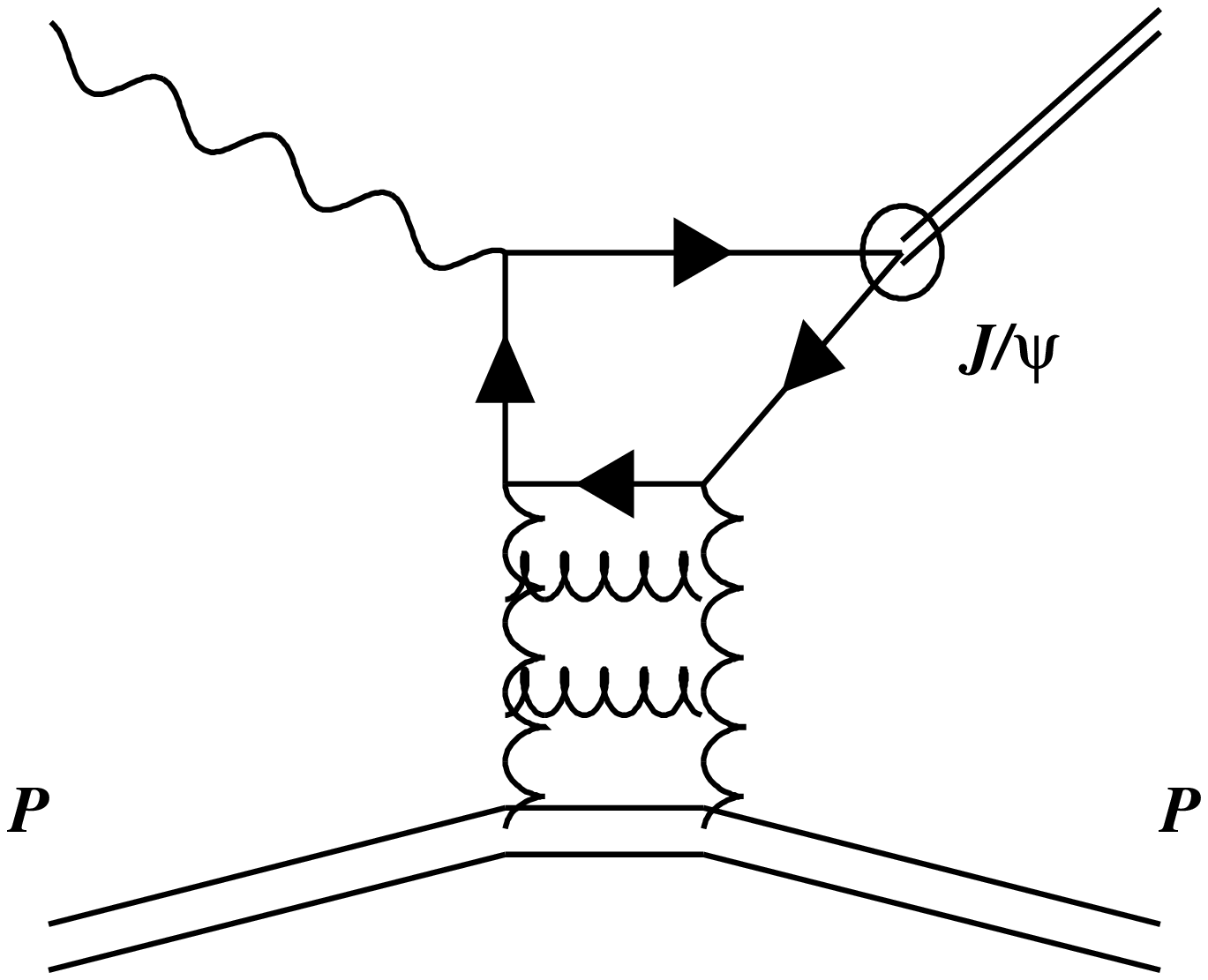,width=6.cm}
 \end{tabular}
\end{center}
\begin{picture}(0.,0.)
\setlength{\unitlength}{0.8cm}
\put(3.5,6.5) {(a)}
\put(10.,6.5) {(b)}
\end{picture}
\vspace{-1.3cm}
\caption{\it 
Leading Feynman diagram for inelastic \jpsi~production (a) and 
a QCD inspired model for elastic \jpsi~production (b) in 
$\gamma$p-collisions.
\label{fig:jpsi}}
\end{figure}
%%%%%%%%%%%%%%%%%%%%%%%%%

 The parton distributions are well constrained for large \xB~where a lot
 of data are available. However, in the small \xB~regime 
 where gluons play a crucial r\^ole, they are not well known.
 HERA has therefore the unique possibility to investigate 
 the distribution of the gluon density in the proton.
 The scaling violation of the proton structure function 
 \ftwo~offers an indirect way to get a handle on the gluon density.
 This measurement was made possible by the extremely successful
 description of \ftwo~by the DGLAP equations and
 has been presented by M. Kasemann at this conference.
 An alternative way is to directly tag the quark box formed
 in a photon-gluon collision 
 either by identifying heavy quarks, e.g. charm via $D^*$ or
 \jpsi~mesons, or by measuring inclusively the 2+1~jet cross section
 at small \xB. This method has, compared to heavy flavour
 production, a much larger cross-section, but has the
 disadvantage that the background from quark initiated
 processes has to be subtracted statistically.
 Moreover, the reach to small values of 
 $\xB_{g/p}$=\xB $(1+\hat{s}/Q^2)$ (see Fig.\ref{fig:feynjets} for notation)
 is restricted, since large 
 invariant jets masses ($\hat{s}>100$\GeVsq) are experimentally required 
 to define clean jets.
  An extraction of the gluon density in leading order using the inclusive 
 2+1~jet rate achieved by the H1 collaboration\cite{h1jetgluon}
 has already been presented on the last 'Physics in Collision'
 conference\cite{goerlach}.

%%%%%%%%%%%%%%%%%%%%%%%%%%%%%%%%%%%%%%%
%\vspace{-9.cm}
\begin{figure}
 \begin{center}
  \begin{tabular}{cc}
  \mbox{ \hspace{-0.7cm}
  \epsfig{figure=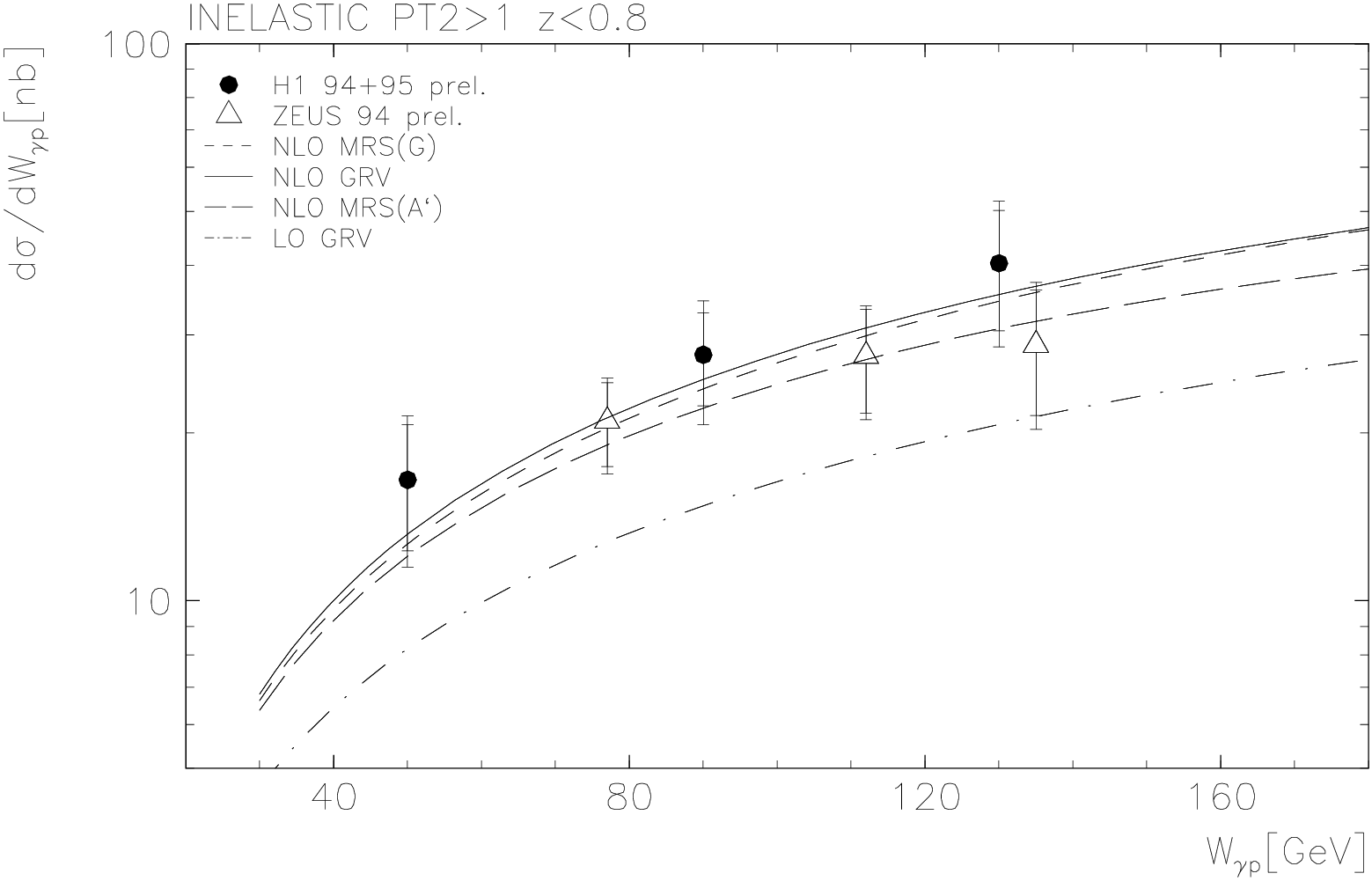,width=4.5cm,angle=90.,
         ,bbllx=32,bblly=805,bburx=532,bbury=30,clip=}
  }
  \mbox{\hspace{6.2cm} 
%  \epsfig{figure=wgpela.huge.ps,width=4.5cm,angle=90.,
%           ,bbllx=20,bblly=748,bburx=501,bbury=61,clip=}
%  }
  \epsfig{figure=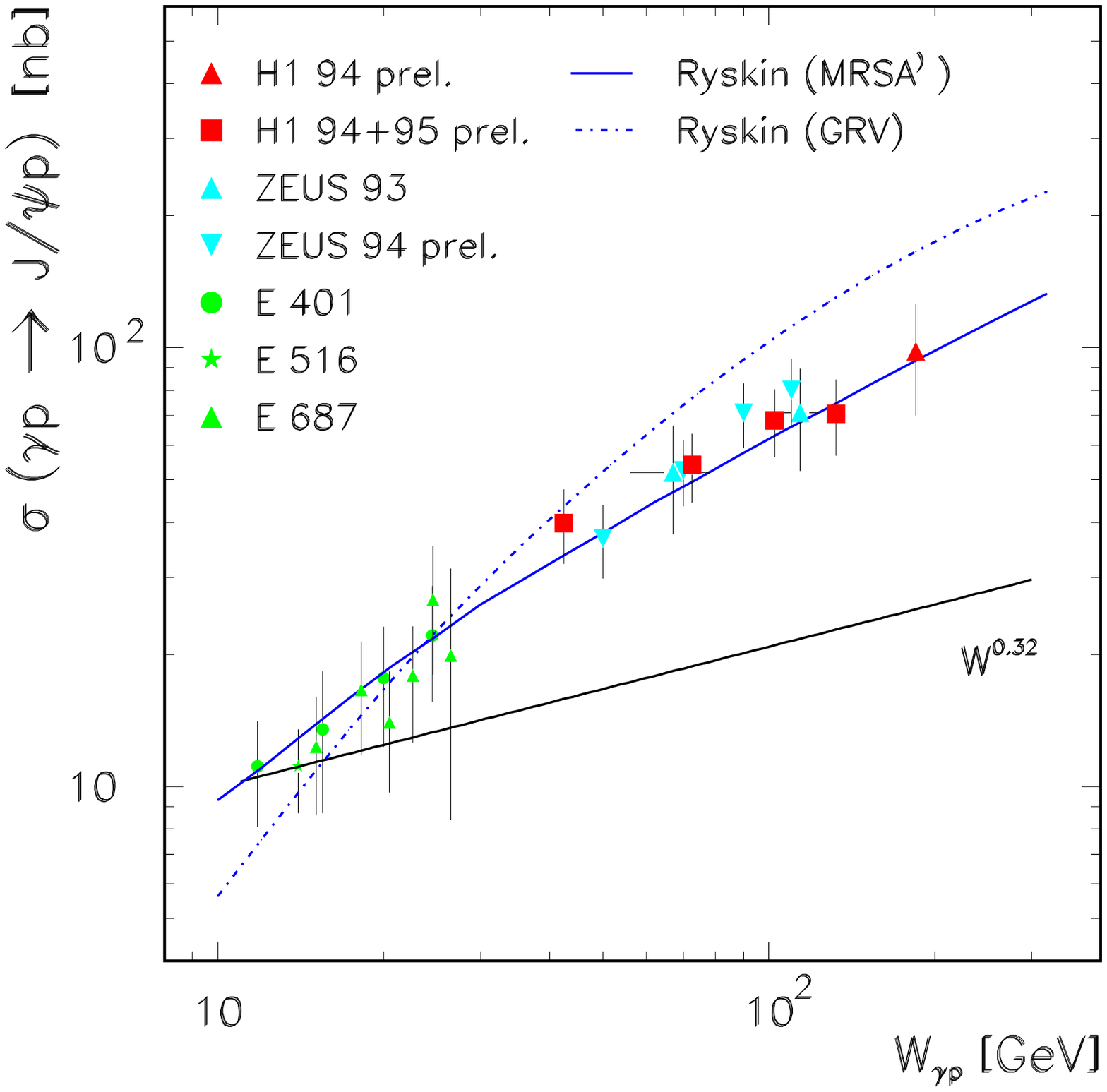,width=6.cm}
}
 \end{tabular}
\end{center}
\begin{picture}(0.,0.)
 \setlength{\unitlength}{0.8cm}
 {\tiny
%  \put(13.5,1.) {$W_{\gamma p}$[{\rm GeV}]}
%  \put(3.,5.) {(a)}
%  \put(11.,5.5) {(b)}
  \put(3.,5.) {(a)}
  \put(14.,6.5) {(b)}
 }
\end{picture}
\vspace{-0.5cm}
\caption{Cross-section for inelastic (a) and elastic (b)
\jpsi~production as function of \Wgp~for HERA and fixed target 
 photoproduction data. 
 Predictions of the Ryskin model for different parameterisations of the gluon
 density are superimposed. The solid line below the data indicates
 a simple parametrisation $\sigma_{\gamma p} \sim \Wgpd$ for 
 $\delta=0.32$.
\label{fig:jpsicross}}
\vspace{-0.5cm}
\end{figure}
%%%%%%%%%%%%%%%%%%%%%%%%%

 Gluon initiated processes can be unambiguously pinned down by
 selecting events where a \jpsi~is produced.
 The leptonic decay modes ($\jpsi \to e^+e^-$ and $\jpsi \to \mu^+\mu^-$)
 of this meson provide a clean experimental signature.
 In the elastic channels only two leptons are found in the
 detector, in the inelastic channel there are
 additional particles (see Fig.~\ref{fig:jpsi}).
 Since the charm quark mass is large enough
 to provide a hard scale, this
 measurement can be performed at very low \Qsq~where
 the cross-section is large.
 Besides the requirement of at least one additional particle in the
 final state, the inelastic channel can be inclusively
 defined using the inelasticity variable
 $z=P \cdot P_{\sjpsi}/(P \cdot q) < 0.8$ where $P$ ($q$)
 is the incoming proton (photon) momentum and $P_{\sjpsi}$ the 
 four-momentum of the \jpsi.
 % z>0.45 (corrected for in result)
 To avoid the phase space region where multiple emissions of gluons become
 more and more important and where a resummation would be required,
 the \pt~of the \jpsi~should exceed $2$\GeV.
 In this region a QCD calculation\cite{jpsiinel} can predict the
 cross-section\cite{jpsiinelcross} 
 of \jpsi~production in next-to-leading order.
% charm quark mass of $1.44$\GeV and \lambdams$=???$.
 The dependence of the cross-section on the photon-proton 
 invariant mass (\Wgp) and the overall
 normalisation is correctly described (see Fig.~\ref{fig:jpsicross}a).
 In this calculation the \jpsi~was assumed to be produced in a
 colour singlet state, no room is left for an additional component
 of a colour octet as required by \pp~data\cite{jpsitev} at high energy.
 However, excluding the region of small \pt, significantly reduces
 the sensitivity to the small-\xB~behaviour 
 of the gluon density as can be seen in 
 Fig.~\ref{fig:jpsicross}a where
 cross-section calculations using various parametrisations
 are compared to data.
 %\input blabla
%%%%%%%%%%%%%%%%%%%%%%%%%%%%%%%%%%%%%%%
\begin{figure}
\vspace{-0.5cm}
\begin{center}
%\mbox{\hspace{-15.cm}
\epsfig{figure=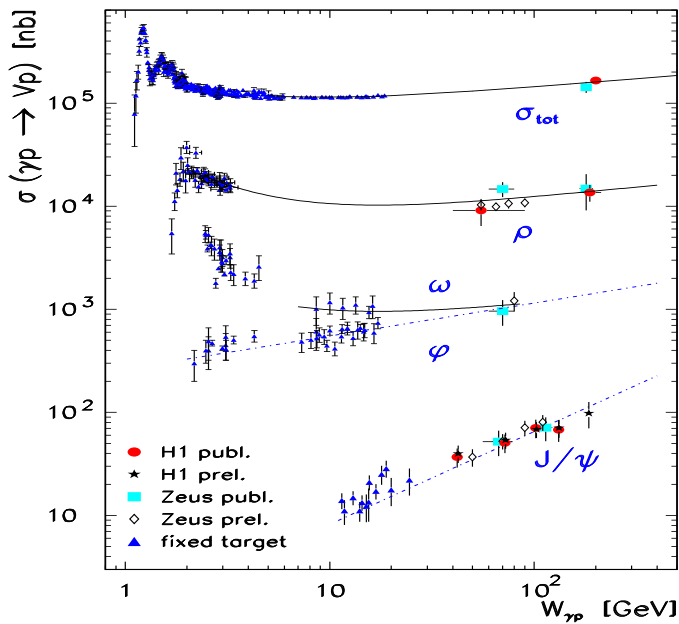,width=10cm,height=9.cm}
%}
\end{center}
%\begin{picture}(0.,0.)
% \setlength{\unitlength}{0.8cm}
% {\Large
%  \put(12.,0.8) {$W_{\gamma p}$}
%  \put(-2.,8.) {$d\sigma/dW_{\gamma p}$}
% }
%\end{picture}
\vspace{-0.5cm}
\caption{Total and vector meson photoproduction cross-section
 as function of \Wgp~measured at HERA and by fixed target
 experiments. Superimposed is a
 simple parametrisation $\sigma_{\gamma p} \sim \Wgpd$ for 
 $\delta=0.32$ (light vector mesons) and $\delta=0.6$ (\jpsi~meson).
\label{fig:gpcross}}
\vspace{-.5cm}
\end{figure}
%%%%%%%%%%%%%%%%%%%%%%%%%

 Previous measurements\cite{h1jpsiold} of elastically produced
 \jpsi~mesons revealed an exponentially falling distribution of the
 momentum transfer between the incoming and outgoing proton
 as expected from diffractive processes and a transverse
 polarisation in agreement with $s$-channel helicity conservation.
 However, the \Wgp~dependence of the elastic cross-section
 \spsiel~measured over a wide range by fixed target
 experiments\cite{jpsifixed} and at HERA\cite{jpsicrossel} 
 exhibits a much steeper rise (see Fig.~\ref{fig:jpsicross}b) than 
 predicted by models based
 on diffractive mechanisms like e.g. the one by Donnachie and
 Landshoff\cite{dola} which assumes a soft interaction in a 
 Regge form where no
 quantum numbers are exchanged. This model relies on the
 observation that any total hadronic cross-section follows
 a power law with the energy available in the center of mass $s$
 like $\sigma \sim s^{\lambda}$ with $\lambda \approx 0.08$.
 If one parameterizes \spsiel~in the region
 $30 < \Wgp < 260$\GeV~as $\spsiel \sim \Wgpd$ one finds
 $\delta=0.6 \pm 0.12$ rather than $\delta=0.32$ as predicted by
 the Regge-inspired model.
 This significant rise can be better described by
 Ryskin's QCD inspired model\cite{ryskin} where the \jpsi~couples
 via two gluons to the proton. To leading order it
 predicts that {\small $\spsiel \sim
 {[\alpha_s(\bar{Q}^2) \; \bar{x} g(\bar{x},\bar{Q}^2) ]}^2 $} 
 where $\bar{Q}^2=1/4 M^2_{\!\sjpsi}$ and
 $\bar{x}=M^2_{\!\sjpsi}/\Wgpsq$.
 The quadratic dependence on \gx, makes the
 elastic channel of \jpsi~photoproduction a sensitive
 discriminator for the slope of the gluon distribution
 $\spsiel \sim {[x \gx]}^2
           \sim x^{-2 \lambda}
           \sim \W^{4\lambda}_{\!\gamma p}/Q^2 $.
 This model, including higher order corrections\cite{ryskinhigh},
 agrees with the data when using the MRSA' parametrisation\cite{mrsap} with
 $\lambda=0.17$.
 The parametrisation of GRV tends to give a steeper \Wgp~dependence. 
 Since phenomenological assumptions
 concerning the two gluon interaction have to be made, the
 \Wgp~dependence of \spsiel, rather than the
 absolute normalization, is a more reliable probe of \gx.
 When comparing to HERA data, the model prediction is therefore
 normalized to fixed target data. A direct extraction of \gx~seems 
 feasible once experimental and theoretical problems are
 better understood.

 The steep rise of \spsiel~with increasing \Wgp~can 
 be contrasted with the moderate
 increase of the total photoproduction cross-section
 and the cross-section of the light vector mesons
 $\rho$, $\omega$ and $\varphi$ which are in excellent agreement
 with model predictions where only soft interactions are
 assumed (see Fig.\ref{fig:gpcross}). 
 The failure of this picture to describe elastic 
 \jpsi~production together with the success of Ryskin's model
 gives additional confidence in the potential of QCD to describe
 processes where the scales involved are still
 surprisingly small.
 The production of light and heavy vector mesons at
 HERA will allow to gain a better understanding of the transition
 from a region governed by soft interactions to a domain where perturbative
 QCD turns on.
 First indications for a similar transition when studying light
 vector meson production with increasing \Qsq~have also been observed at
 HERA\cite{zeuslight}.

\section*{Deep-inelastic scattering at small \xB}
\label{sec:lowx}
%%%%%%%%%%%%%%%%%%%%%%%%%%%%%%
\begin{figure}
%\hspace{-1cm}
\begin{center}
\begin{tabular}{cc}
 \epsfig{figure=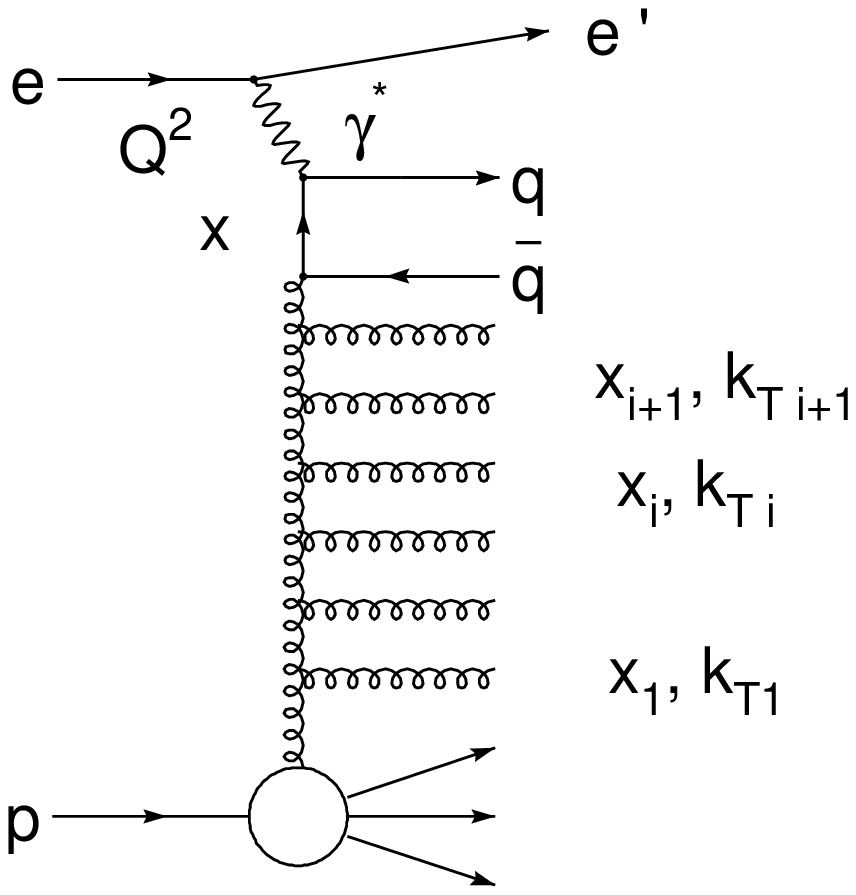,width=6cm}
% \vspace{-1cm}
 \mbox{\hspace{-4cm}
 \epsfig{figure=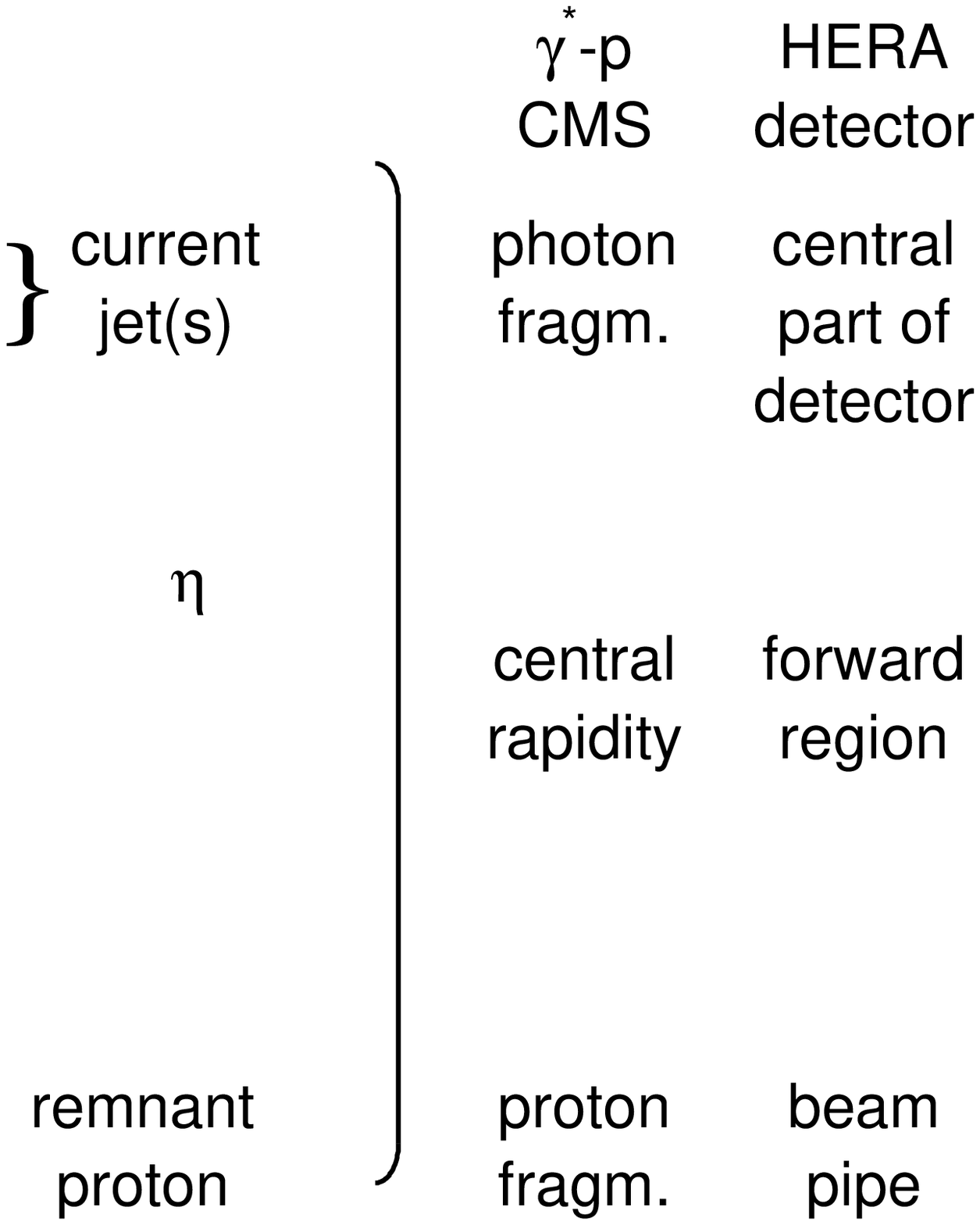,width=6cm}
 }
\end{tabular}
\end{center}
%        bbllx=401,bblly=729,bburx=137,bbury=323
%        ,width=8cm,angle=270,clip=}
\vspace{-0.5cm}
\caption{\it Diagram of an \ep~collision at low \xB.
\label{fig:gluonladder}}
\vspace{-0.3cm}
\end{figure}
%%%%%%%%%%%%%%%%%%%%%%%%%

 At low \xB~the simple picture of deep-inelastic scattering
 as a process where a virtual photon interacts
 instantaneously with a point-like parton 
 freely moving in the proton has to be modified.
 Since the probability that a parton radiates becomes
 increasingly high at low \xB, the parton struck by the photon
 originates most likely from a cascade initiated
 by a parton with high longitudinal momentum.    
 The phase space for gluon radiation
 is enlarged with decreasing \xB. %and the rapidity
% distance between the photon and the proton grows logarithmically.
 Such an interaction is illustrated in Fig.\ref{fig:gluonladder}.
 The gluon is the driving force behind the parton cascade. 
 The transitions $g \to gg$ and $g\to q\bar{q}$ at each point
 in the ladder can be approximated by the DGLAP equations.
 By resumming the LO and NLO ${(\alpha_s \ln{Q^2})}^n$ terms they predict
 the \Qsq~evolution of a parton known to be point-like at some given
 scale $Q_0$ to the region where the interaction with the photon
 takes place. They describe the change in the parton
 density with varying spatial resolution ($\lambda \sim 1/Q$) 
 of the probe. One of the assumptions\footnote{For a 
 digestible introduction to low-\xB~physics see A. Martin\cite{martin}.}
 necessary to derive the DGLAP equations is a strongly ordered
 configuration in the parton virtualities  along the ladder 
 connecting the soft proton constituents to the hard subprocess. 
 This leads to a suppression of the available
 phase space for gluon radiation. 
 
 When the ${(\alpha_s \ln{1/x})}^n$ terms become large, they
 have to be taken into account, e.g. by the resummation
 accomplished by the BFKL equations\cite{BFKL}.
 In a physical gauge, these terms correspond
 to an $n$-rung ladder diagram in which gluon emissions
 are ordered in longitudinal momentum. The strong ordering of
 the transverse momenta is replaced by a diffusion pattern
 as one proceeds along the gluon chain.
% $\gx=\int^Q2 dk_t^2/k_t^2 f(x,k_t^2)$ with
% $ f(\xB,k_t^2)= \xB^{-\lambda_L} 
%   \exp{(-\ln^2{(k_t^2/A)}/B\ln{1/\xB})}
%  \lambda=(3\als/pi) 4 \ln{2}.
% $
 The BFKL equations describe how a
 particular high momentum parton in the proton is dressed by
 a cloud of gluons at low \xB~localized in a fixed transverse 
 spatial region of the proton. 
% If the available rapidity range
% ($\eta \sim \ln{1/\xB}$) becomes so large that many low-\xB~gluons 
% coexists in the same transverse spatial region, these
% gluons will no longer act as free partons, but will start to interact
% among themselves. 
 At very low \xB~it is expected that many gluons coexist and will
 no longer act as free partons, but interact with each other. 
 This 'saturation' regime is characterized by an equilibrium 
 of gluon emission and absorption.
 A new QCD regime is reached where individual parton-parton 
 interactions are weak, but where the field strength becomes
 - due to the number of partons - so strong that perturbation 
 theory is not reliable.
% The growth of the parton density
% with decreasing \xB~at fixed \Qsq~is therefore expected to slow down
% and eventually to stop.
 It is not clear if in the HERA regime such effects could
 be observed.

 Experimentally, it is interesting to measure observables 
 sensitive to the underlying parton dynamics.% in such a 
% complex object like the proton.  
 Studying the probe after the interaction is the classical
 approach to get information on the structure of the target.
 The measurement of the proton structure function $F_2$
 has the benefit that it can be directly compared to analytical QCD 
 calculations.
 The rise of \ftwo~with decreasing \xB~observed at HERA\cite{F2} 
 appears to be well described by the DGLAP equations.
 Even at the lowest values of \xB~the BFKL terms
 are not needed to obtain a satisfactory description.
% However structure functions based on BFKL are also compatible
% with data\cite{f2bfkl}.

%
%%%%%%%%%%%%%%%%%%%%%%%%%%%%%%%%%%%%%%%
\begin{figure}
\vspace{-0.2cm}
%\begin{center}
%% \setlength{\unitlength}{0.8cm}
% \put(0.,-0.5) {(a)}
% \put(6.,-0.5) {(b)}
% \end{center}
\mbox{ \hspace{-1.cm}
 \epsfig{figure=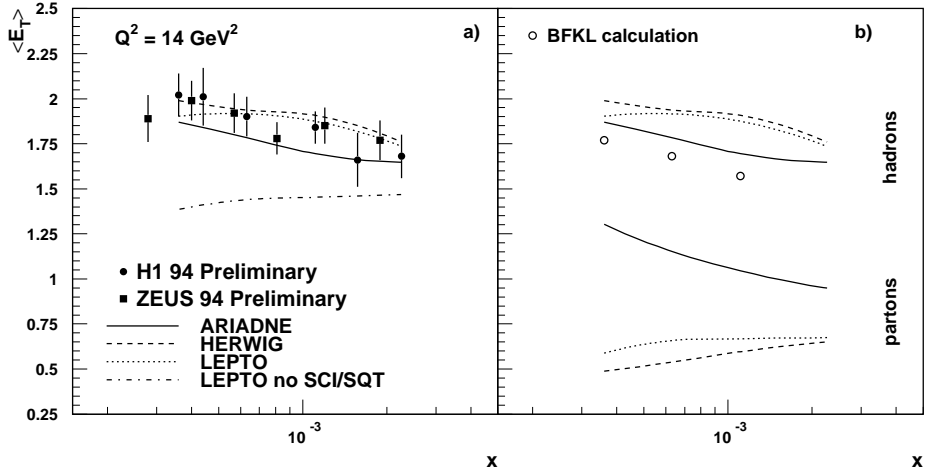,width=13cm}
 }
\vspace{-0.5cm}
\caption{Mean transverse energy 
 in $-0.5 < \eta^* < 0.5$ measured by H1 and ZEUS (closed symbols)
 in the cms for $Q^2=14$~\GeVsq~as function of \xB.
 Superimposed are QCD model predictions (lines) for hadrons (a) and partons (b)
 by QCD models and an analytic BFKL calculation (open circles) carried
 out on parton level only.
\label{fig:etmean}
}
\end{figure}
%%%%%%%%%%%%%%%%%%%%%%%%%%%%%%%%%%%%%%

%%%%%%%%%%%%%%%%%%%%%%%%%%%%%%%%%%%%%%%
\begin{figure}
\vspace{-1.cm}
% \mbox{ \hspace{-1.cm}
 \epsfig{figure=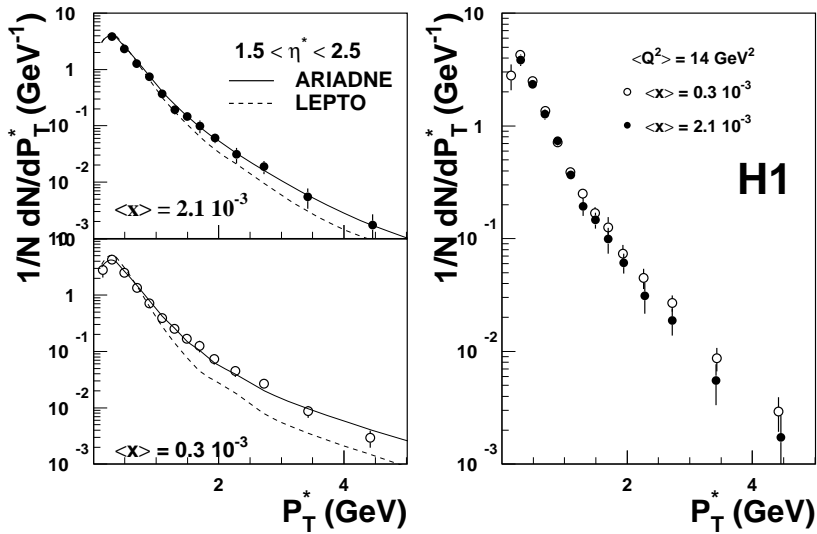,width=12.cm}
% }
%
\vspace{-0.7cm}
\caption{Transverse momentum distribution of single charged particles
 in $1.5 < \eta^* < 2.5$ measured by H1 in the cms 
 for fixed $Q^2=14$\GeVsq~and two \xB~bins.
 Superimposed as lines are QCD model predictions.
\label{fig:etpt}
}
\vspace{-0.5cm}
\end{figure}
%%%%%%%%%%%%%%%%%%%%%%%%%%%%%%%%%%%%%%

%
 However, $F_2$ might be too inclusive to reveal the potentially small
 resummation effects. 
 A better indicator might be provided by observables
 based on the hadronic final state emerging from the parton
 cascade, since they can be chosen such 
 that the differences between the two complementary approximations
 are enhanced. However, predictions have then mainly to be
 based on Monte Carlo event generators 
 simulating the detailed properties of the hadronic final 
 state by including the hard subprocess, QCD radiation effects
 and the transition of partons to hadrons.

  The LEPTO Monte Carlo\cite{lepto} incorporates the QCD matrix element
 to first order in \als~and approximately  accounts for
 higher order parton emissions by the concept of parton 
 showers\cite{partonshower} based on the DGLAP equations (MEPS model).
 The non-perturbative transition of the partonic final state to the
 observable hadrons is modeled by the
 LUND string model\cite{lund} as incorporated in JETSET\cite{jetset}. 
 ARIADNE\cite{ariadne} differs only in the treatment of
 the hard subprocess and the QCD shower evolution.
 Gluon emissions are treated by the colour dipole model\cite{cdm}
 assuming a chain of independent radiating
 dipoles spanned by colour connected partons. The first emission
 in the cascade is corrected by the matrix element to first 
 order\cite{ariadneme}. 
 A phenomenological suppression of gluons with short
 wavelengths (having high transverse momenta) only resolving
 part of the extended colour charge distribution of the proton 
 remnant has been introduced for DIS. This leads to 
 a more pronounced population of partons towards the proton 
 remnant than observed in LEPTO where the phase space is controlled
 by the strong ordering of the gluon virtualities and the ratio
 of the parton density function at each branching point.
 It has been argued that the partonic state as encountered
 in ARIADNE is more closely related to the one expected from a
 BFKL scenario\cite{muellerbfkl,rat96}.
 %Despite all these differences,
 %both Monte Carlo models are able to give a consitient picture of the
 % event topology and provide a fair description of basic
 % event properties.
 Both Monte Carlo models are able to give a consistent picture of the
 event topology and provide a fair description of basic
 event properties\cite{carli96,herawsmctun}.  
 
 Using the hadronic final state,
 the presence of parton activity can be experimentally probed 
 from the region close to the photon quark interaction
 down to (pseudo-)rapidities defined as 
 $\eta^* = - \ln{\tan{(\theta/2}})$ where 
 $\theta$ is the polar angle between a particle and the proton 
 direction\footnote{
 The direction of the proton points along the negative $z$-axis in the cms.}. 
 In the cms $\eta^* \approx -1$ 
 can be reached beyond which the acceptance of the two main HERA detectors 
 ends (see also Fig.\ref{fig:gluonladder}). 
 An inclusive measure is the mean transverse energy (\av{\et})
 in the central region $-0.5 < \eta^* < 0.5$.
 Due to the increasing phase space with decreasing \xB, 
 gluons are more abundant and \av{\et} is expected to rise. 
 This behaviour is seen in the data\cite{etflow93,etflow96,zeusflow96}
 as is illustrated in Fig.\ref{fig:etmean}a where the \av{\et} is shown
 for fixed $\Qsq=14$\GeVsq.
 Both QCD models are able to describe the data. 
 LEPTO, however, has to produce $60 - 80$\%
 of the \av{\et} during the hadronization phase to compensate for
 the different \xB~dependence of the \av{\et} seen at the parton level
 as is illustrated in Fig.\ref{fig:etmean}. 
 This could only be achieved by introducing two new 
 non-perturbative production mechanisms for \av{\et}.
 In events where
 a sea-quark is involved in the hard subprocess, its partner is not
 - as in previous versions - simply rearranged to a meson or a
 baryon within the proton remnant, but is used to stretch a string
 to one of the quarks remaining untouched in the proton (SQT).
 The second new ingredient is the assumption of a soft colour 
 interaction\cite{sci}(SCI) which changes only the colour configuration of the 
 partonic system while leaving the colour field of the proton.
 This colour rotation offers, by producing a colour singlet in
 the final state, a possible explanation of the rapidity gap events
 observed at HERA and at the same time (due to longer strings
 caused by the colour rearrangement) the transverse energy needed
 to match the data.
 Without these two new features a constant amount of \av{\et} is
 generated during the hadronisation and the data are not described
 (see Fig.\ref{fig:etmean}).
 Although ARIADNE uses the same hadronisation model as LEPTO, 
 a constant amount of only 30-40\% of the \av{\et} is produced
 during hadronisation.
 An analytical BFKL calculation\cite{bfklet}, only carried out for partons, 
 predicts the same \xB~dependence as seen in the data, but lies
 significantly above the corresponding ARIADNE curve.  

 The hard tail of the \pt~distribution of single charged particles
 has been suggested\cite{kuhlen} as a better tool to disentangle
 non-perturbative hadronisation effects from parton activity
 as described by perturbative QCD. This observable is more
 directly related to hard parton emissions and cannot be
 mimicked by a cumulative effect from many soft particle emissions 
 during hadronization. The measured steeply falling \pt~distribution 
 in the interval $1.5 <\eta^*<2.5$ gets 
 harder for decreasing \xB~(see Fig.\ref{fig:etpt}).
 While at large \xB~LEPTO and ARIADNE are both able to describe the
 data, LEPTO significantly falls below the data at low \xB.
 This indicates that more partons are produced than expected
 from models with suppressed parton radiation
 based on the DGLAP parton showers.
 Similar, but less conclusive, results\cite{etflow96} 
 have been obtained by using the inclusive transverse energy distribution
 in $- 0.5 <\eta^*< 0.5$.

%\input blabla
%%%%%%%%%%%%%%%%%%%%%%%%%%%%%%%%%%%%%%%
\begin{figure}
\vspace{-0.5cm}
\begin{center}
\mbox{\hspace{-1.cm}
\epsfig{figure=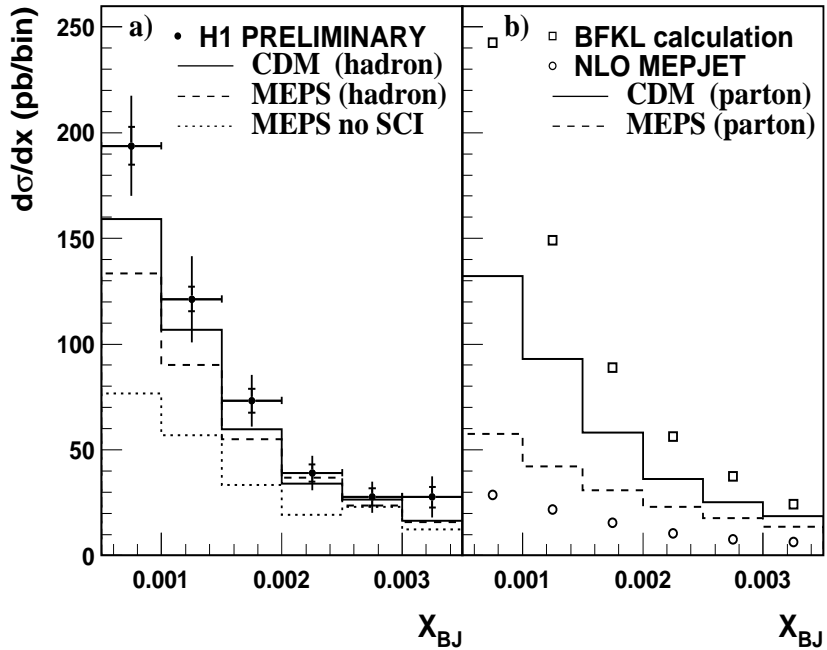,width=12.cm,height=9.cm}
%         ,bbllx=85,bblly=641,bburx=512,bbury=236,clip=}
}
\end{center}
%\begin{picture}(0.,0.)
% \setlength{\unitlength}{0.8cm}
% {\Large
%% \put(1.,5.5) {$\sqrt{s} \approx 300$~{\rm GeV}}
% \put(10.,6.8) {$x= \frac{Q^2}{p \cdot \gamma} $}
% \put(10.,5.2) {$W^2 \approx Q^2 \frac{1-x}{x}$}
% }
%\end{picture}
%\vspace{9.cm}
\caption{\it Forward jet cross-section as function of 
\xB~measured by H1. Superimposed are QCD model predictions for
hadrons (a) and partons (b). An analytic BFKL calculation
(open squares) and a NLO QCD calculation (open circles)
are superimposed.  
\label{fig:fjets}}
\vspace{-0.3cm}
\end{figure}
%%%%%%%%%%%%%%%%%%%%%%%%%

%%%%%%%%%%%%%%%%%%%%%%%%%%%%%%%%%%%%%%%
\begin{figure}
%
%\setlength{\unitlength}{0.8cm}
% \put(0.,-0.5) {(a)}
% \put(6.,-0.5) {(b)}
 \mbox{ \hspace{-0.5cm}
 \epsfig{figure=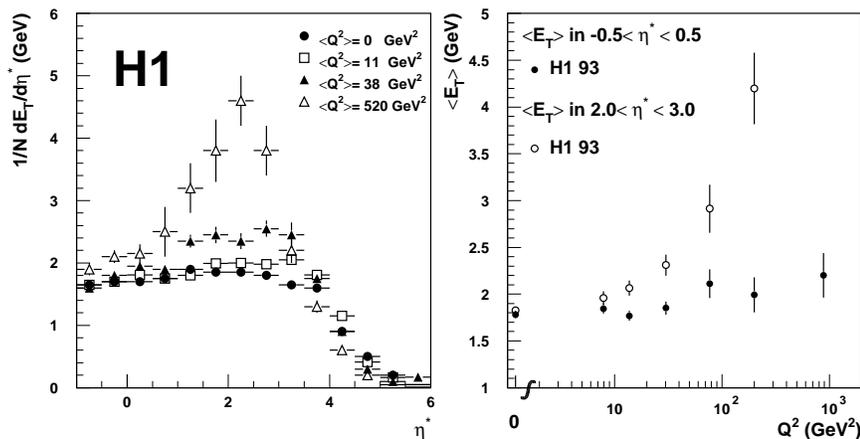,width=12cm}
 }
\vspace{-0.5cm}
\caption{Energy flow at fixed $W=180$~\GeVsq~ in the hadronic center of
 mass for different \Qsq~bins ranging from $0$ to $520$\GeVsq~ and
 mean transverse energy for $-0.5 < \eta^* <0.5$ and $2 < \eta^* <3$
 in function of \Qsq.
\label{fig:rosti}}
\end{figure}
%%%%%%%%%%%%%%%%%%%%%%%%%%%%%%%%%%%%%%

 Another signature designed to enhance BFKL effects is the production
 of 'forward jets'~\cite{fjets} characterized by a squared transverse
 momentum ($k_T^2$) of the order of the photon virtuality 
 \Qsq~and $x_{\rm jet}=E_{\rm jet}/E_p$ as large and \xB~as small
 as kinematically possible.   
 $E_{\rm jet}$ ($E_p$) denotes the energy of the forward jet (proton).
 The first requirement suppresses the strongly ordered DGLAP evolution.
 Asking for large $x_{\rm jet}/\xB$,
 the forward jet can be separated by a large rapidity
 interval from the struck quark such that the phase space
 of parton emission between the two is amplified.  
 For this kinematical configuration the $\alpha_s \ln{x_{\rm jet}/\xB}$ 
 terms are expected to become so large that their resummation should
 lead to a sizable increase of the forward jet cross-section. 
 %Furthermore, for jets near the proton
 %remnant with  the ambiguity of the cross-section
 %on the parton density function should be small.

A fast rise of the cross-section for forward jets defined by  
 $k_T> 3.5$\GeV, $ 0.5 < k_T^2/\Qsq< 2$ and $x_{\rm jet}>0.035$
 with decreasing \xB~is seen
 in the H1 data\cite{h1fjets} corrected to hadron level
 (see Fig.\ref{fig:fjets}a). ARIADNE falls slightly below
 the data, but shows the same trend towards small \xB. 
 Hadronisation effects are well below $20$\% such that
 the same \xB~dependence is also seen on parton level.
 While LEPTO predicts - as expected - a similar moderate increase
 of the forward jet cross-section as a full NLO QCD
 calculation\cite{mirkesfjet} at the parton level, it is also able
 to describe the data by producing a large fraction of forward
 jets (up to $80$\%) in the hadronisation phase. Without assuming
 soft colour interaction the hadronisation corrections
 are similar to the ones obtained in ARIADNE, but then the
 data are not described (dotted line in Fig.\ref{fig:fjets}a).
 An analytical BFKL calculation\cite{fjetbfkl} exhibits a much
 faster rise with $1/\xB$ compared to the ARIADNE result on parton
 level. However, the authors point out that several effects 
 which might lower the prediction have not been taken into account.

  \section*{Deep-inelastic scattering as hadron-hadron scattering}
 Strictly speaking, the physical picture of DIS as an interaction
 of a photon with a parton freely moving in the proton is only
 valid if the proton is very fast. Only in this reference frame
 is the interaction time of the virtual photon with the parton 
 short compared to the interactions regularly taking place 
 in the proton such that the photon instantaneously probes
 the proton content. In the reference frame
 where the proton is at rest and the photon is fast, it is
 more appropriate to imagine that the photon fluctuates into
 a quark anti-quark pair before interacting with the proton.
 If the fluctuation time is long compared to the interaction time, 
 a parton cascade
 can develop and a complicated object is formed which 
 strongly interacts with matter. 
% However, if the mass of the
% pair\footnote{Note, $\xB=\Qsq/W$ and M_{\rm pair}\approx \Qsq.} 
% is small the parton cascade can not be calculated by pert. 
% QCD and some phenomenological form factor have to used. 
 Using the uncertainty principle the fluctuation 
 time\cite{dokshitzer} 
 can be calculated to $\tau_{\gamma}=1/(m_p \xB)$ where $m_p$ is the 
 proton mass. For a typical \xB~value of $10^{-3}$, the distance
 before the photon interacts with the target
 is $c \tau_{\gamma} \approx 200$~{\rm fm} which is large compared
 to the proton radius of about $1$~{\rm fm}. 

%%%%%%%%%%%%%%%%%%%%%%%%%%%%%%%%%%%%%%%
\begin{figure}
\vspace{-2.cm}
\begin{center}
%\mbox{\hspace{2.cm}
\epsfig{figure=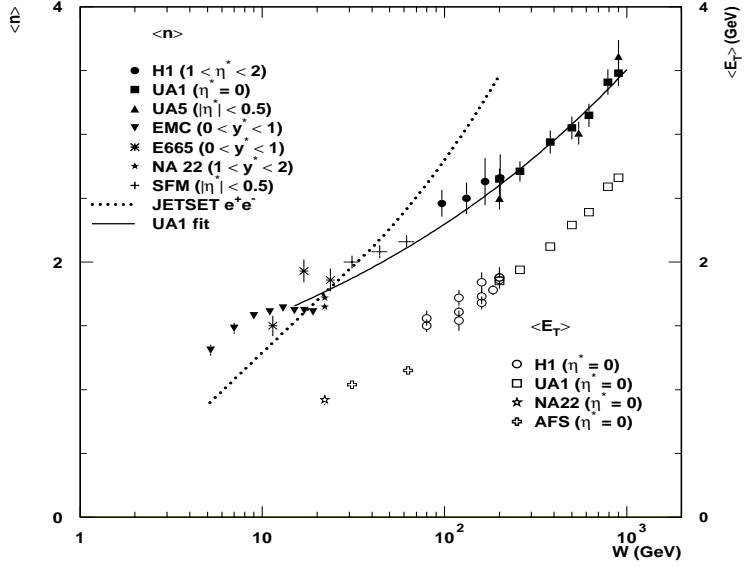,width=10.cm,height=8.cm,
         ,bbllx=4,bblly=24,bburx=500,bbury=491,clip=}
%}
\end{center}
\caption{Mean multiplicity of charged particles in a given (pseudo-)rapidity 
region $y$($\eta$) and \av{\et}~in a central rapidity region measured
in \ep~and \pp~collsions as function of $W$. 
The solid (dotted) line indicates the behaviour of the mean charged particle
multiplicity in \pp~(\ee)~collsions. 
\label{fig:mult}}
\vspace{-0.5cm}
\end{figure}
%%%%%%%%%%%%%%%%%%%%%%%%%
 
 If one chooses suitable observables, one can therefore find in
 DIS at low-\xB~characteristic properties of hadron-hadron collisions.
 Changing the photon virtuality from $\approx 0$ to
 about $520$\GeVsq, the transverse energy flow\cite{eflowfixed}
 in the hadronic cms for fixed $W=180$\GeV~varies
 only in the region where the hard subprocess has taken place 
 (see Fig.\ref{fig:rosti}a). 
% Since at high \Qsq~the quark anti-quark
% pair is close together, their \pt~must be large according to the
% uncertainty principle. 
 The \av{\et} in $2 < \eta^* < 3$ indeed grows
 with \Qsq~as is depicted in Fig.\ref{fig:rosti}b. Surprisingly,
 this rise only starts at about $10$\GeVsq. The measured \av{\et} 
 produced in collisions with on-shell photons, which cannot 
 be treated perturbatively, has the same magnitude as observed
 in a high \Qsq~region where first calculations have been very
 promising. 
 In the central rapidity region of $-0.5 < \eta^* < 0.5$ however,
 \av{\et} is only little influenced by the virtuality of the photon.
% The impact of the photon virtuality is lost after evolving
% over a distance of $\Delta\eta^* \approx 1.5$ like it is typical for 
% a strong interaction known to be short-range.
 The impact of the photon virtuality penetrates only
  $\Delta\eta^* \approx 1.5$. This is typical of the short range
 correlation in inelastic hadron-hadron collisions.
 When comparing the \av{\et} to hadron-hadron collisions,
 it scales with the center of mass energy of the collision independent
 of the nature of the incoming particles.
 This is shown in Fig.\ref{fig:mult}. It is interesting to note
 that also the mean charged particle multiplicity
 exhibits a similar dependence on $W$.
 Also here the DIS data from HERA\cite{h1mult} interpolate
 - despite the slightly different rapidity regions -
 between previous fixed target data and hadron-hadron collisions
 at much higher energy (see Fig.\ref{fig:mult}).
 This suggests a universal behaviour of the dynamics in the
 central rapidity region.
 In \ee~collisions a steeper dependence
 on $W$ is seen for these rapidity regions 
 (dotted line in Fig.\ref{fig:mult}).
 
 Since DIS data reveal some properties of soft hadron-hadron interactions
 there is some hope that the understanding of DIS at low \xB~by
 perturbative QCD can give a handle on the understanding of
 the underlying dynamics of strong interactions.
 If the first attempts to understand in perturbative QCD
 some low-\xB~HERA data are successful, a tool might become available
 to provide a universal description of strongly interacting 
 particles from an perturbative theory of asymptotically free
 partons at short distances to the theory of the confined
 hadrons at long distances.

\section*{Acknowledgments}
It is a pleasure to thank my colleagues from the H1 and ZEUS
collaboration for providing me with the latest data and
information. I have profited from many stimulating discussions 
with M.~Kuhlen and E.~DeWolf and I gratefully appreciate their 
careful reading of the manuscript. 
I deeply acknowledge the support of the physics coordinators
of the H1 and ZEUS collaboration, J.~Dainton and R.~Nania, 
and would like to thank them for their helpful comments.
 
% Kuhlen
% DeWolf
% Dainton
% would like to thank
% have profited from many discussions
%It is a pleasure to thank
%We gratefully appreciate the outcome 
% We are deeply indebted for his regular se
%
\section*{References}
\end{document}